\documentclass[useAMS,usenatbib]{mn2e}
\usepackage{graphicx}
\usepackage{epsfig}
\usepackage{amsmath}
\usepackage{amssymb}
\usepackage{xcolor}
\usepackage{ulem}
\usepackage{appendix}
\usepackage{multirow}

\newcommand{\be}{\begin{equation}}
\newcommand{\beq}{\begin{equation}}
\newcommand{\ba}{\begin{eqnarray}}
\newcommand{\ee}{\end{equation}}
\newcommand{\eeq}{\end{equation}}
\newcommand{\ea}{\end{eqnarray}}

\def\lsim{~\rlap{$<$}{\lower 1.0ex\hbox{$\sim$}}}

\def\gsim{~\rlap{$>$}{\lower 1.0ex\hbox{$\sim$}}}

\title[LBG clustering at $z\sim4$]{The clustering and halo occupation distribution of Lyman-break galaxies at $z\sim4$}	
\author[Jaehong~Park et al.]
      {Jaehong~Park$^{1}$\thanks{jaehongp@student.unimelb.edu.au}, Han-Seek~Kim$^{1}$\thanks{hansikk@unimelb.edu.au}, J. Stuart B.~Wyithe$^{1}$, C. G. Lacey$^{2}$,\\ \vspace{0.3mm}\\
       {\LARGE {\rm C. M. Baugh$^{2}$, R. L. Barone-Nugent$^{1}$, M. Trenti$^{1}$ and R. J. Bouwens$^{3,4}$}}\\
       $^1$School of Physics, The University of Melbourne, Parkville, VIC 3010, Australia\\
       $^2$Institute for Computational Cosmology, Department of Physics, University of Durham, South Road, Durham DH1 3LE, UK\\
       $^3$Leiden Observatory, Leiden University, NL-2300 RA Leiden, Netherlands\\
       $^4$UCO/Lick Observatory, University of California, Santa Cruz, CA 95064, USA
	}
\date{}

\pagerange{\pageref{firstpage}--\pageref{lastpage}}
\pubyear{2015} 

\begin{document}

\maketitle

\label{firstpage}


\begin{abstract}
We investigate the clustering of Lyman-break galaxies (LBGs) at $z\sim4$. Using the hierarchical galaxy formation model GALFORM, we predict, for the first time using a semi-analytical model with feedback from active galactic nuclei (AGN), the angular correlation function (ACF) of LBGs and find agreement within $3\,\sigma$ with new measurements of the ACF from surveys including the Hubble eXtreme Deep Field (XDF) and CANDELS field. Our simulations confirm the conclusion reached using independent models that although the predicted ACFs reproduce the trend of increased clustering with luminosity, the dependence is less strong than observed. We find that for the detection limits of the XDF field central LBGs at $z\sim 4$ predominantly reside in haloes of mass $\sim 10^{11}-10^{12}h^{-1}M_{\rm \odot}$ and that satellites reside in larger haloes of mass $\sim 10^{12}-10^{13}h^{-1}M_{\rm \odot}$. The model predicts fewer bright satellite LBGs at $z\sim4$ than is inferred from measurements of the ACF at small scales. By analysing the halo occupation distribution (HOD) predicted by the model, we find evidence that AGN feedback affects the HOD of central LBGs in massive haloes. This is a new high-redshift test of this important feedback mechanism. We investigate the effect of photometric errors in the observations on the ACF predictions. We find that the observational uncertainty in the galaxy luminosity reduces the clustering amplitude and that this effect increases towards faint galaxies, particularly on small scales. To compare properties of model with observed LBGs this uncertainty must be considered.
\noindent 
\end{abstract}



\section{Introduction}

In hierarchical galaxy formation models, galaxies form inside dark matter haloes. The growth of dark matter haloes can be successfully described by analytical models \citep[e.g.][]{MW96,Cooray2002} and by N-body simulations \citep[e.g.][]{Springel2005,MII2009}. The assembly of galaxies and the evolution of properties such as luminosity and stellar mass can be calculated 
starting from the growth of dark matter haloes \citep{Lacey2011,Tacchella2013,Mason2015}. However, the galaxy formation process itself remains poorly understood \citep[e.g.][]{Baugh2006, Benson2010,Schaye2015}. One way to investigate the astrophysical connection between dark matter haloes and galaxies is by comparing models with observational estimates of galaxy clustering. 

At high redshift, Lyman-break galaxies (LBGs) are the most extensively studied sources \citep[e.g.][]{Giavalisco2002}. LBGs are star-forming galaxies detected by a spectral feature which arises because the rest-frame far-UV emission is absorbed (below $1216\, {\rm \AA}$) by neutral hydrogen. Since the original work of \cite{Steidel&Hamilton1993} and \cite{Steidel1996} at $z\sim3$, this technique has been extended to detect galaxies up to $z\sim10$ \citep{Oesch2012,Bradley2012,McLure2013,Zitrin2014,Duncan2014,Bouwens2015}.  Luminosity functions in the rest-frame UV have been measured from LBG observations, along with properties such as their star formation rates, stellar masses, and dust extinction \citep{Gonzalez2011,Smit2012,Lee2012,Wilkins2013}. 

Measurements of the angular correlation function (hereafter ACF) of galaxies probe the connection between dark matter haloes and galaxies. On large scales the observed ACFs can be approximated by a power-law, $w(\theta)=A_{w}\theta^{-\beta}$, where $A_w$ is the angular correlation amplitude and $\beta$ is the correlation slope. Over the past decade the ACF of LBGs has been measured at high redshifts, $z \sim 3.5-6$  \citep{Ouchi2005,Lee2006,Cooray2006,Kashikawa2006b,McLure2009,Hildebrandt2009} and recently out to $z\sim7.2$ \citep{Rob2014}.  These observations show an enhanced clustering amplitude compared to a power law on small scales. In order to understand the form of the ACF, the two-point correlation function can be decomposed into one- and two-halo terms (see \citealp{Cooray2002} for a review).  On scales larger than the typical halo virial radius, the clustering amplitude is dominated by the contribution from galaxy pairs in separate haloes. This is called the two-halo term. On the other hand, galaxy pairs inside the same halo contribute to the clustering amplitude on small scales. This is called the one-halo term and is interpreted as arising from central-satellite and satellite-satellite galaxy pairs. 

The clustering can be modelled using the halo occupation distribution (HOD), which describes the number of galaxies per halo as a function of a host halo mass \citep{Peacock&Smith2000,Seljak2000,BWS2002,Berlind&Weinberg2002}. This is typically expressed as the mean number of galaxies per halo, with some assumption about the width of the distribution \citep{Benson2000}. In the local Universe, the HOD has been studied using N-body simulations, smoothed particle hydrodynamics (SPH) simulations and semi-analytical calculations \citep{Benson2000,Berlind2003, Kravtsov2004, Zheng2005}, and it has been inferred from observations \citep[e.g.][]{Zehavi2011, Torre2013}. At high redshifts, HOD studies have mainly focused upon empirical approaches based on observations \citep{Hamana2004,Hamana2006,Cooray2006,Lee2006,Hildebrandt2009,Lee2009}

There have been a number of empirical studies which have attempted to interpret galaxy clustering at high redshifts ($z \gtrsim 3$) (e.g. \citealp{Rob2014} and references therein). However, less attention has been given to the predictions of a priori theoretical models. \cite{Mo1998} studied the formation of disc galaxies in hierarchical clustering models, while \cite{Governato1998}, \cite{Baugh1998,Baugh1999} and \cite{Kauffmann1999} used semi-analytical models to interpret galaxy formation and evolution up to $z\sim3$. \cite{Wechsler2001} investigated the clustering properties of LBGs at $z\sim3$ to probe the nature of LBGs using N-body simulations combined with semi-analytical models. \cite{Kravtsov2004} studied the clustering properties up to $z=5$ by analysing the HOD and the halo two-point correlation function of subhaloes using N-body simulations, but they did not include galaxy formation physics. Based on the Subaru Deep Field, \cite{Kashikawa2006b} compared the observationally measured clustering of LBGs at $z\sim4$ and 5 with the prediction of mock LBGs generated using a semi-analytical model combined with an N-body simulation. \cite{Jose2013} also investigated galaxy clustering at high redshift with a semi-empirical calculation. \cite{Wechsler2001} calculated the number of model galaxy pairs using a simple assumption that every dark matter halo above a mass threshold has one visible LBG, and found that this is not consistent with observations on small scales. Recent studies show that models which have multiple LBGs in massive haloes can explain the clustering amplitude on small scales \citep{Kravtsov2004,Kashikawa2006b,Jose2013}.

Recently, the number of high-redshift galaxies observed has increased dramatically. \cite{Bouwens2015} identified LBGs up to $z\sim10$ in a combined survey field consisting of the Hubble eXtreme Deep Field (XDF) and CANDELS fields, which are the deepest existing surveys. \cite{Rob2014} studied the clustering properties of these samples by measuring the ACF. The measurements from different fields allow us to assess the sample variance in the ACF. Motivated by this observational advance, we revisit the predictions for the clustering of LBGs from current models of galaxy formation. To investigate the clustering properties of LBGs at $z\sim4$ we use the hierarchical galaxy formation model GALFORM \citep{Cole2000}. In particular, we use a recent version of the model described in \cite{Lagos2012}. This is the first test of the predictions for the angular clustering of high redshift galaxies using a semi-analytical model which includes AGN feedback. In this study, we predict the ACF of LBGs selected in the model and compare the results with the ACF measured from observations. By comparing the predicted and measured ACFs, we can analyse the clustering properties of central and satellite galaxies, and study the implications for the form of the LBG HOD.

We begin in Section~\ref{model} by describing the GALFORM model. In Section~\ref{LBGs}, we present the methodology used to select LBGs and to compute ACFs in the model. We describe the ACF measured from the observations in Section~\ref{Observed_ACF}. Then, we present the predictions for the ACF and compare with observations in Section~\ref{Comparisons_with_observations}. We show the dependence of clustering on luminosity in Section~\ref{Clustering_dependence}. In Section~\ref{HOD} we analyse the model HOD. We summarise in Section~\ref{Summary}. The Appendix discusses the effect on the clustering signal caused by the photometric scatter.

Throughout the paper we use apparent magnitudes in the observer frame on the AB system. Where we refer to the UV magnitude, this corresponds to the rest-frame $1500{\rm \AA}$ AB magnitude.

\section{The model}\label{model}
In this section we summarise the model used in this study. In Section~\ref{GALFORM}, we briefly review GALFORM. Then we describe photoionisation feedback in Section~\ref{Feedback}.

%
%
\subsection{The GALFORM galaxy formation model}\label{GALFORM}

The semi-analytical model GALFORM computes the formation and evolution of galaxy properties within a hierarchical structure formation framework  (see \citealp{Cole2000} and \citealp{Lacey2015} for a comprehensive overview of GALFORM and \citealp{Baugh2006} for a review of semi-analytical models). We implement GALFORM within the Millennium-II dark matter simulation \citep{MII2009}. The particle mass is 6.89$\times$10$^{6}$$h^{-1}{\rm M_{\odot}}$ and we use haloes with 20 particles or more (the minimum halo mass corresponds to $\sim 1.4\times 10^{8}h^{-1}{\rm M_{\odot}}$) although for comparison to observed LBGs we consider only galaxies located in haloes with mass above  $\sim 2.8\times 10^{9}h^{-1}{\rm M_{\odot}}$ which ensures that various properties of the dark matter halos 
can be measured robustly \citep{Trenti2010}. The simulation box has a side length $L=100h^{-1}$Mpc. We consider galaxies with a baryonic mass (cold gas plus stars) greater than $10^{6}h^{-1}{\rm M_{\odot}}$ in the output of the semi-analytic model. This resolution is suitable for haloes which host the faint galaxies in the XDF and CANDELS survey fields. The Millennium-II simulation has a cosmology defined by fractional total and baryonic mass, and dark energy densities of $\Omega_{\rm m}=0.25$, $\Omega_{\rm b}=0.045$ and $\Omega_{\Lambda}$=0.75, a dimensionless Hubble constant of $h$=0.73, and a power spectrum normalisation of $\sigma_{8}$=0.9. Note that we base our study on the halo merger trees described in \cite{Jiang2014} which are designed for the purposes of semi-analytic modelling in GALFORM (see also \citealp{Merson2013}). 

\begin{figure*}
\begin{center}
\includegraphics[width=18cm]{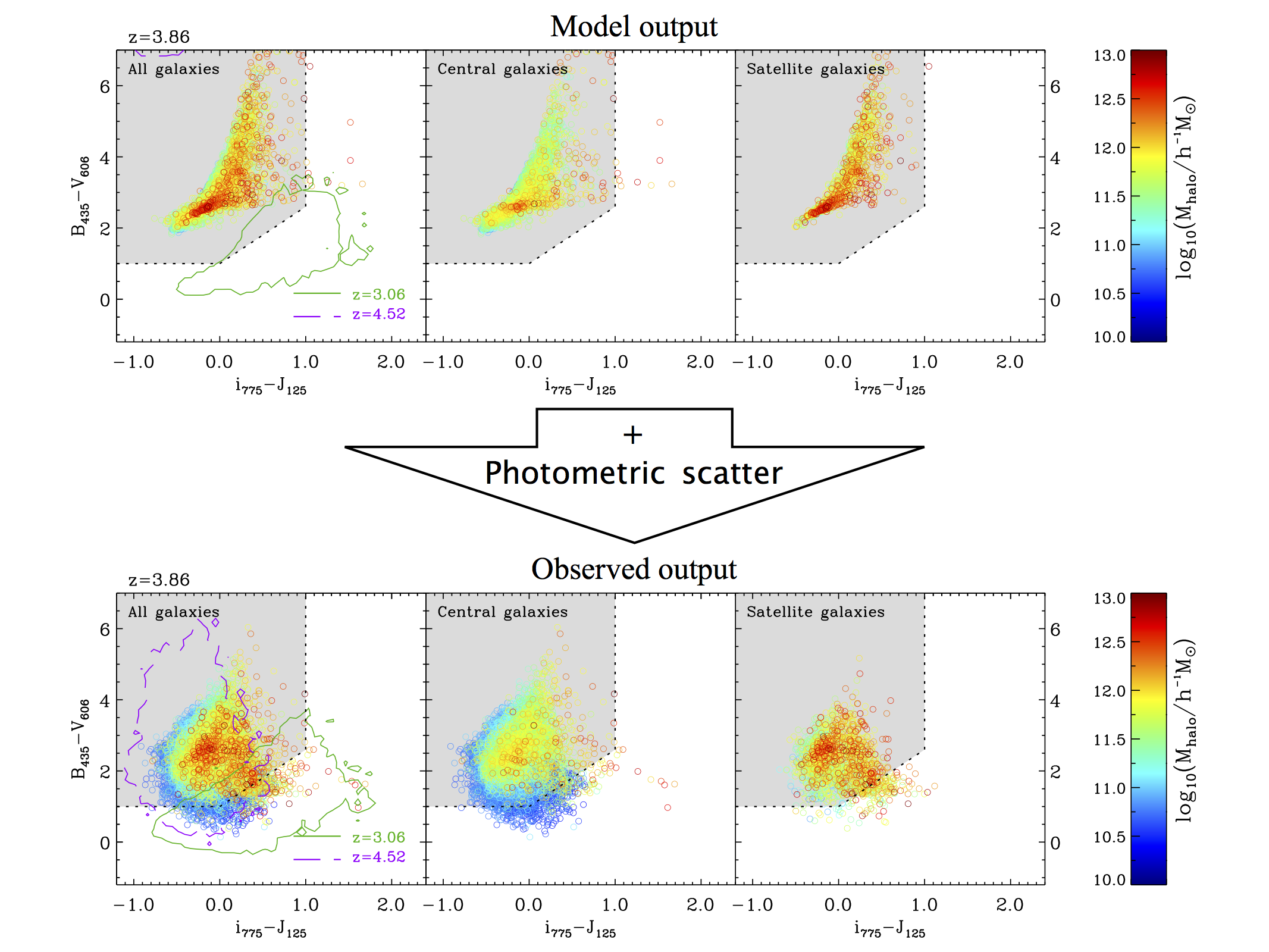}
\end{center}
\vspace{-3mm}
\caption{Colour-colour diagram showing model galaxies and the selection criteria used at $z=3.86$. Top and bottom panels show the intrinsic simulated galaxy colours (i.e. without photometric uncertainty) and colours including photometric scatter for XDF field, respectively. Green and violet contours in the left panels enclose 99.7 per cent of the galaxies at $z=3.06$ and 4.52. Grey shaded regions and dotted lines represent the selection criteria area adopted for LBGs by \protect\cite{Bouwens2015}. Different colour symbols correspond to the host halo mass as indicated by the colour bar. The panels show galaxies brighter than the XDF flux limit. The left panels show all galaxies, the middle panels show central galaxies and the right panels show satellite galaxies.}
\label{Fig1}
\end{figure*}

Here, we use the \cite{Lagos2012} version of GALFORM described in \cite{Kim2012a}. The \cite{Lagos2012} model uses the improved star formation treatment implemented by \cite{Lagos2011} which split the interstellar medium in galaxies into HI and $\rm H_2$ phases using the observationally motivated relation of \cite{Blitz2006}, with $\rm H_2$ providing the fuel for star formation. This is a key difference from previous models which assumed that the star formation law applied to all of the cold gas in galaxies (cf. \citealp{Cole2000, Bower2006, Baugh2005}).

%
%
\subsection{Feedback processes}\label{Feedback}
Feedback processes play a key role in galaxy formation. GALFORM includes three main feedback processes; Supernova (SN) feedback which suppresses the formation of galaxies within small dark matter haloes, AGN feedback which suppresses the formation of galaxies in massive haloes by shutting down gas cooling, and photoionisation feedback \citep{Cole2000,Benson2002,Benson2003,Bower2006,Kim2011,Kim2012b,Lacey2015}. Here, we briefly explain the implementation of photoionisation feedback, which differs from the standard implementation, and is designed to account for the patchy nature of reionisation.

\subsubsection{Photoionisation feedback}\label{Photo_feedback}
A strong ionising background leads to several physical effects such as the suppression of cooling by photo-heating \citep{Efstathiou1992}, higher IGM gas pressure \citep{Gnedin2000} and photoheating \citep{Hoeft2006,Okamoto2008}. As a result, star formation is suppressed within ionised regions of the IGM during reionisation \citep[see][]{Crain2009}, which may result in self-regulation of the reionisation process \citep{iliev2007}. GALFORM includes a simple prescription for this process in which the cooling of halo gas is suppressed in haloes with circular velocity below a value $V_{\rm cut}$ when the IGM becomes globally ionised at a particular redshift $z < z_{\rm cut}$ \citep{Benson2002}. In the standard implementation of GALFORM, the onset redshift is assumed to be $z_{\rm cut}=10$. 

Instead of a constant $z_{\rm cut}$, we use the prescription described in \cite{Kim2012a} to take into account that reionisation proceeds at different rates in different locations. \cite{Kim2012a} divide galaxies from the model into cells of small volume, and calculate the number of photons produced by galaxies in the cell that enter the IGM and trigger reionisation. They then calculate the ionisation fraction in each cell and find HII regions which are defined as a region that has an ionisation fraction that is larger than unity. The star formation is suppressed only in the HII regions. In this model, reionisation starts in some patches at $z\sim12$, and is assumed to end at $z=6$.

\begin{figure*}
\begin{center}
\includegraphics[width=18.5cm]{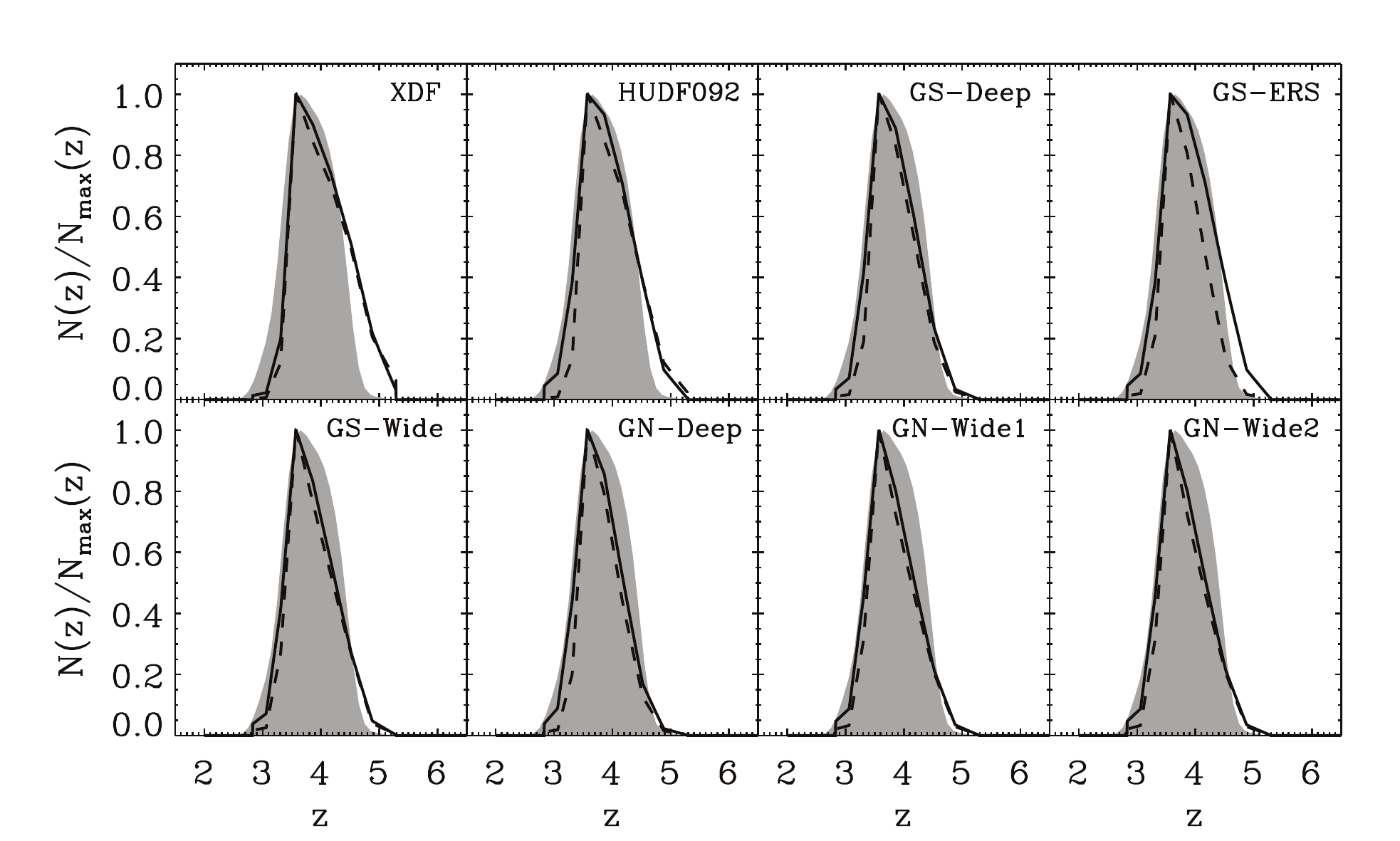}
\end{center}
\vspace{-3mm}
\caption{
Redshift distributions of selected LBGs for different flux limits corresponding to the observational flux limits of the XDF and CANDELS survey fields as labelled in each panel. The redshift distributions are normalised to have a maximum value of unity. Dashed and solid lines represent, respectively, the redshift distribution of Intrinsic-LBGs and Obs-LBGs. Shaded regions represent the redshift distribution of observed LBGs estimated by \protect\cite{Bouwens2015}.
}
\label{Fig2}
\end{figure*}

\section{Lyman-break galaxies selected in the model}
\label{LBGs}
In this section we describe how we select Lyman-break galaxies (LBGs) from the model and how we compute the angular correlation function (ACF) of these galaxies.
\subsection{Selecting Lyman-break galaxies}
\label{LBG_select}
In a previous study using GALFORM, \cite{Lacey2011} showed that the \cite{Baugh2005} model predicts the observed rest-frame UV luminosity function over the redshift range $z=3-10$. Using the same model, \cite{Gonzalez-Perez2013} studied the rest-frame UV colours of LBGs in the redshift range $z= 2.5-10$, confirming that various colour selection criteria \citep{Steidel1995,Bouwens2012,Wilkins2011,Lorenzoni2011} are effective at selecting model galaxies in the desired redshift ranges.

We select LBGs using a similar method to that described in \cite{Gonzalez-Perez2013}, but using the colour selection criteria from \cite{Bouwens2015}. To identify candidate sources, \cite{Bouwens2015} developed  selection criteria using deep optical and near-IR observations ($i_{775}$ and $J_{125}$). For $z\sim4$ (B-dropout technique), the criteria are given by
\begin{gather}\label{colour-selection}
(B_{435}-V_{606} >1) \ \wedge \ (i_{775}-J_{125} <1) \ \wedge \\
       (B_{435}-V_{606}) >1.6(i_{775}-J_{125})+1, \nonumber
\end{gather}
where $\wedge$ represents the logical AND symbol. 

We calculate magnitudes for model galaxies in each band starting from the star formation history computed for each galaxy (see e.g. \citealp{Gonzalez-Perez2013}). The magnitudes of galaxies are attenuated since dust grains in the inter-stellar medium absorb photons emitted from the stars. Within GALFORM, this attenuation is computed using the radiative transfer model of \cite{Ferrara1999} (see \citealp{Cole2000,Lacey2011}). Another factor affecting the magnitudes is absorption by neutral hydrogen. Photons emitted by galaxies are absorbed by the intergalactic medium, including those from the Lyman continuum and Lyman series. To apply this effect to the apparent magnitudes, GALFORM uses the transmission formulation proposed by \cite{Madau1995}. We have checked that replacing the Madau transmission with that from \cite{Meiksin2006} has little impact on our results.

We take into account the observational flux limits given in \cite{Bouwens2015}. LBGs are observed in combined survey fields consisting of the Hubble eXtreme Deep Field (XDF) and CANDELS survey. Since the individual survey fields have different flux limits, we select LBG samples for each survey field. Table~\ref{Table:Flux_limit} shows the flux limits corresponding to the individual survey fields expressed as apparent magnitudes corresponding to a 5\,$\sigma$ depth. 

We also take into account the effect on measured galaxy colours caused by a non-detection and by a photometric scatter, which can have a significant impact on faint galaxies. Observations have a flux detection limit and a flux uncertainty. In cases of a non-detection, \cite{Bouwens2015} set the flux in the drop-out band ($B_{435}$) to be equal to the 1\,$\sigma$ flux limit to calculate a $(B_{435}-V_{606})$ colour. In this case the measured colours denote a lower limit on the true colour. The relation between flux at the 5\,$\sigma$ detection limit and the 1\,$\sigma$ detection limit is $m_{1\sigma}=m_{5\sigma}+2.5{\rm log_{10}}(5/1)$, where $m_{5\sigma}$ is the 5\,$\sigma$ detection limit listed in Table \ref{Table:Flux_limit}. To mimic the observations, we therefore replace the predicted magnitudes in the $B_{435}$ band with the 1\,$\sigma$ detection limit when the magnitude in the $B_{435}$ is fainter than this limit. We also apply the photometric scatter to the simulated magnitudes using
\begin{equation}\label{Mag}
m' = -2.5{\rm log_{10}}(10^{-0.4\times m} + {\rm noise}),
\end{equation}
where the noise term represents a random Gaussian flux uncertainty with a mean value of zero. A 1\,$\sigma$ noise level is obtained from the 5\,$\sigma$ magnitude limit listed in Table~\ref{Table:Flux_limit}, using ${\rm noise_{1\sigma}}=10^{(-0.4\times m_{5\sigma})}/5$. We detect model galaxies using the stated 5\,$\sigma$ magnitude limits following these modifications, and denote the model LBGs selected including the effects on the colours caused by non-detections and the photometric scatter as `Obs-LBGs'. Since the photometric scatter is decided by the magnitude limits, the Obs-LBGs sample depends on which survey field is selected. We denote the model LBGs selected using intrinsic colours, i.e. without any photometric scatter, as `Intrinsic-LBGs'.

\cite{Bouwens2015} estimated the redshift distribution of observed LBGs using Monte-Carlo simulations. They found that the distribution of redshift  corresponds to a Gaussian with a central redshift of 3.8 and standard deviation of 0.38. To model this redshift distribution, we apply the colour selection criteria to galaxy catalogues in sequential redshift snapshots from the simulation between $z=2.83$ and 5.29 ($z=2.83$, 3.06, 3.31, 3.57, 3.86, 4.17, 4.52, 4.88 and 5.29).

\begin{table}
\begin{center}
\caption{
         Flux limits and areas of survey fields. Each magnitude limit is quoted as a 5\,$\sigma$ depth \protect\citep{Bouwens2015}. 
         }
\begin{tabular} {cccccc}
 \\
 \hline\\[-3.0mm]
 Field                  &   Area [${\rm arcmin^{2}}$]   &    $B_{435}$     &     $V_{606}$       &    $i_{775}$         &            $J_{125}$             \\[0.5mm] \hline\\[-2.5mm]
 XDF                   &          4.7                               &        29.6           &          30.0            &            29.8         &             29.3                     \\[0.5mm]
 HUDF092          &          4.7                               &        28.3           &           29.3            &           28.8         &              28.9                     \\[0.5mm]
 GS-Deep           &          64.5                             &        27.7           &           28.0            &            27.5         &            27.8                        \\[0.5mm]
 GS-ERS            &          40.5                             &        27.5           &           27.7            &            27.2         &             27.6                       \\[0.5mm] 
 GS-Wide           &            34.2                            &       27.7           &           28.0            &             27.5        &              27.1                       \\[0.5mm] 
 GN-Deep           &          62.9                             &       27.5            &           27.7            &             27.3        &              27.7                      \\[0.5mm] 
 GN-Wide\,1\,\&\,2      &   60.9                            &       27.5            &           27.7            &             27.2        &             26.8                        \\[0.5mm] 
  \hline
\end{tabular}
\label{Table:Flux_limit}
\end{center}
\end{table}

Fig.~\ref{Fig1} shows the colours of model galaxies correspond to the selection criteria at $z=3.86$, along with the colour distributions of galaxies at $z=3.06$ and 4.52. The top panels show the colours of Intrinsic-LBGs. At the target redshift of $z\sim3.8$ almost all galaxies are selected as LBGs. This is to be expected, since galaxies at high redshift have ongoing star-formation and so are bright in the rest-frame UV, and the $i_{775}-J_{125}$ colour straddles the Lyman-break, which is present in every galaxy. The colour distribution shifts toward a larger value of $(B_{435}-V_{606})$ at $z=4.52$. On the other hand, most galaxies are outside the selection regions at $z=3.06$. The bottom panels show the colours of Obs-LBGs. This shows a similar trend to the Intrinsic-LBGs, but the colour distribution has more scatter. Compared with the $(B_{435}-V_{606})$ colours of Intrinsic-LBGs the $(B_{435}-V_{606})$ colours of Obs-LBGs do not show very large values. This is because the magnitude computed from the simulation before applying photometric scatter does not have a flux limit in the drop-out band. Overall, this is consistent with the result of \cite{Gonzalez-Perez2013}. Specifically, the selection criteria of \cite{Bouwens2015} successfully exclude model galaxies which lie outside the target redshift range. Although a small number of intrinsic-LBGs and a few per cent of Obs-LBGs at the target redshift are excluded by the selection criteria, the impact on the clustering is negligible.
 
Fig.~\ref{Fig2} shows the predicted redshift distributions, $N(z)$, of selected model LBGs for flux limits corresponding to those of the XDF and CANDELS fields. The redshift distributions plotted in Fig.~\ref{Fig2} are normalised to a maximum value of unity. The $N(z)$ for Intrinsic-LBGs and Obs-LBGs are comparable, with the Obs-LBGs $N(z)$ being slightly wider. The predicted distributions in two of the deep fields (XDF and HUDF092) show more high redshift galaxies than the other fields. This is because the two deep fields have deeper flux limits and so contain more faint galaxies at high-redshift. The observed $N(z)$ from \cite{Bouwens2015} is shown for comparison. Taking into account the fact that \cite{Bouwens2015} estimated $N(z)$ by combining all survey fields, the predicted redshift distributions are in good agreement with observations. On the whole, the predicted redshift distributions are consistent with the observed ones, providing some reassurance that the colour selection criteria work for the model galaxies.  


\subsection{The luminosity function}
\label{LF}

Luminosity functions from observed samples are measured using an effective volume, $V_{\rm eff}$. The effective volume can be written
\begin{equation}\label{V_eff}
V_{\rm eff} = V_{\rm tot}\, p(m,z),
\end{equation}
where $V_{\rm tot}$ is the total survey volume and $p(m,z)$ is the probability of selecting a source of magnitude $m$ at redshift $z$ as a LBG, and takes into account the completeness and selection function of the observations (see e.g. \citealp{Bouwens2015,Finkelstein2015}). In simulations, we consider snapshots in the redshift range $2.83 \leq z \leq 5.29$ for comparison with data, and define the selection function at a given redshift as
\begin{equation}\label{completeness}
p(m) = \sum_{i}^{N}\frac{n_{{\rm sel},i}}{n_{{\rm tot},i}},
\end{equation}
where $i$ is the snapshot number, $N$ is the number of snapshots, $n_{\rm sel}$ is the number of selected LBGs, and $n_{\rm tot}$ is the total number of galaxies in the magnitude bin. Then, the effective volume in each magnitude bin is 
\begin{equation}\label{V_eff-sim}
V_{\rm eff} = \sum_{i}^{N}V_{\rm sim}\, p(m,z_i),
\end{equation}
where $V_{\rm sim}$ is the simulation box size.

\begin{figure}
\begin{center}
\includegraphics[width=8.6cm]{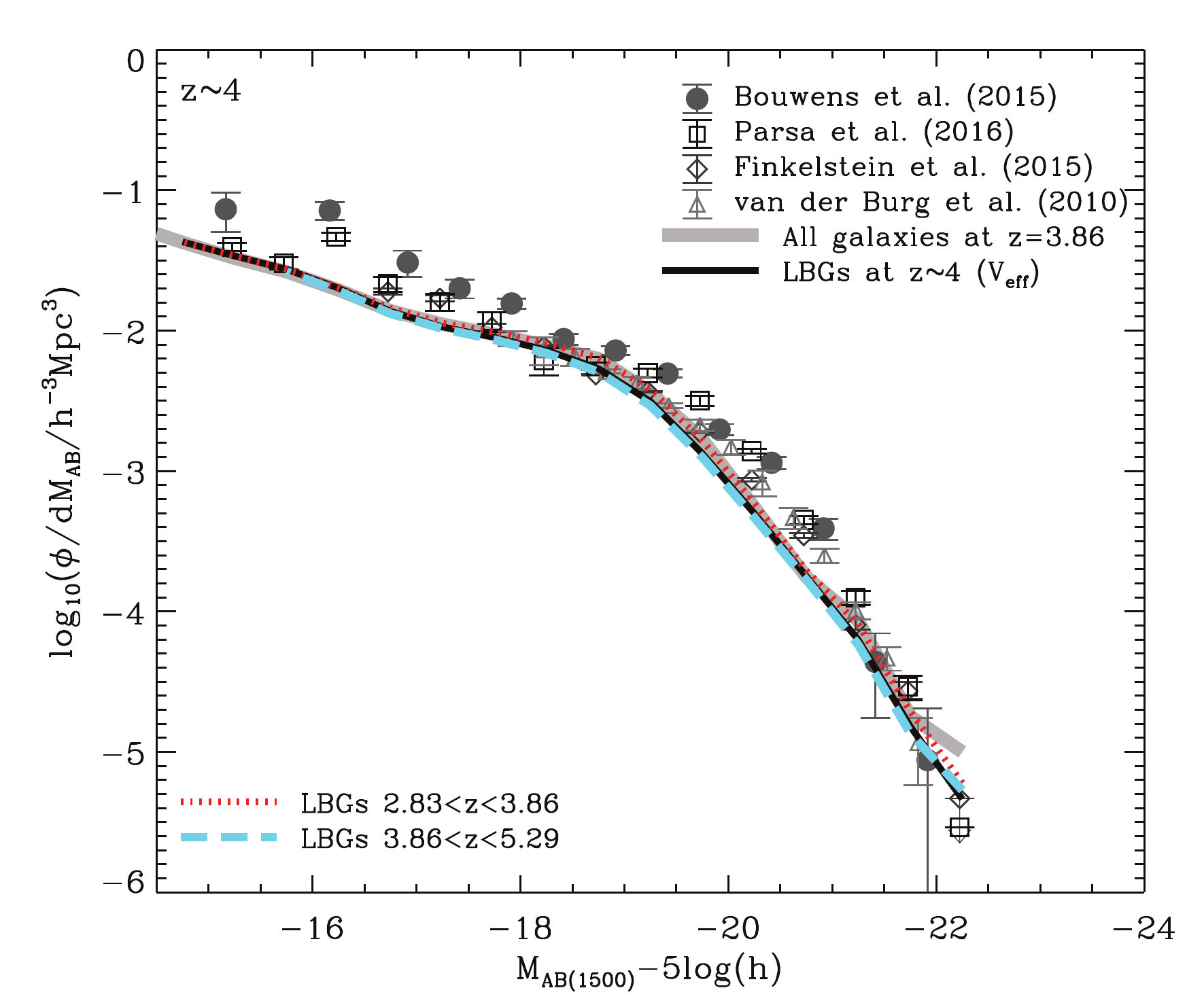}
\end{center}
\vspace{-3mm}
\caption{The predicted rest-frame UV luminosity function with observed luminosity functions. Thick grey line represents the predicted luminosity function using all galaxies in the snapshot at $z=3.86$. Solid line represents the predicted luminosity function of selected LBGs in the range of redshift distribution, $2.83\leq z \leq5.29$. Dotted and long-dashed line represent the predicted luminosity functions of selected LBGs in low ($2.83\leq z \leq3.86$) and high  redshift $3.86\leq z \leq5.29$, respectively. The legend indicates observed luminosity functions from different surveys \citep{VanDerBurg2010, Bouwens2015,Finkelstein2015,Parsa2016}
}
\label{Fig3}
\end{figure}

Fig.~\ref{Fig3} shows the resulting prediction for the luminosity function. We find that the predicted luminosity function in the redshift range $3.86 \leq z \leq 5.28$ stops at $M_{\rm AB(1500)}-5{\rm log}(h) \sim -16$. On the other hand, the predicted luminosity function in the redshift range $2.83 \leq z \leq 3.86$ reaches to the faintest luminosity on the luminosity function measured from observations. This implies that the amplitude of the luminosity function at the faint end is due mainly to galaxies at low redshift $z < 3.86$. This is because the absolute magnitude of a galaxy at low redshift is fainter than at high redshift for a fixed apparent magnitude. We note that two predicted luminosity functions at $2.83 \leq z \leq 3.86$ and $3.86 \leq z \leq 5.29$ show similar amplitudes. Since most selected model LBGs consist of galaxies in the redshift range $3.5 \lesssim z \lesssim 4.5$ and the redshift evolution of the luminosity function over this redshift range is not significant, the two predicted luminosity functions are similar to one another. We also find that across most of the luminosity range the luminosity function predicted over the redshift distribution is consistent with the luminosity function predicted in the snapshot at $z=3.86$. This implies that the luminosity function at the target redshift is representative of the observed luminosity function, even if the observed redshift range is broad.

When comparing the predicted luminosity function with that of \cite{Bouwens2015} we find that the amplitude of the faint end ($M_{\rm AB(1500)}-5{\rm log}(h) \sim -16$) and bright end ($M_{\rm AB(1500)}-5{\rm log}(h) \sim -20$) deviate from the observed luminosity function by $3\,\sigma$. This discrepancy may arise from inaccurate modelling of various physical processes such as feedback (SN and AGN) and dust extinction. However, different studies find slightly different luminosity functions \citep{VanDerBurg2010,Bouwens2015,Finkelstein2015,Parsa2016}. As shown in Fig.~\ref{Fig3}, we find that the predicted luminosity function is comparable with observations.

%
\subsection{Modeling the Angular Correlation Function}
\label{ACF}
In this section, we compute the angular correlation function (ACF) of simulated LBGs. We first calculate the three dimensional two-point correlation function in the model in each redshift slice. From the density field $\rho({\bf x})$, the two-point correlation function is defined as
\begin{equation}\label{xir}
\xi({\bf r}) = \left< \delta({\bf x}) \delta({\bf x+r}) \right >,
\end{equation}
where $\delta({\bf x}) = \rho({\bf x})/\bar{\rho} -1$ with $\bar{\rho}$ being the mean density. In the model, which has a periodic volume, we compute the two-point correlation function using the excess probability, compared to a random distribution, of finding a pair of galaxies with separation $r$ to $r+\delta r$,
\begin{equation}\label{xip}
1+\xi(r)=\frac{DD}{\bar{n}^2}\frac{1}{V{\rm d}V},
\end{equation}
where $DD$ is the number of pairs of galaxies, $\bar{n}$ is the mean galaxy density, $V$ is the volume of the simulation box and ${\rm d}V$ is the differential volume between $r$ and $r+\delta r$. 

From the two-point correlation function, we then calculate the ACF using Limber's equation \citep{Limber1954},   given by
\begin{equation}\label{Limber}
w(\theta)=\frac{2\int_{0}^{\infty}\left[ N(z)\right]^{2}/R_{\rm H}(z) \left( \int_{0}^{2r} {\rm d}u \ \xi(r_{12},z) \right) {\rm d}z} {\left[ \int_{0}^{\infty} N(z) {\rm d}z\right]^2},
\end{equation}
where $N(z)$ is the redshift distribution of selected galaxies and $R_{\rm H}(z)$ is the Hubble radius. For comoving distances $r_1$ and $r_2$ to a pair of galaxies, we denote $u=r_1 - r_2$, $r_{12}=\sqrt{u^2 + r^2\theta^2}$ and $r=(r_1+r_2)/2$, using the small angle approximation. In order to integrate the two-point correlation function to large scales beyond which the model cannot make accurate predictions due to the finite simulation box size, we assume that the two-point correlation function follows the power-law, $\xi(r) = (r/r_0)^{-\gamma}$, where $r_0$ is the correlation length and $\gamma$ is the correlation slope.  
For this we obtain the best fitting parameters $r_0$ and $\gamma$ to the predicted two-point correlation function in the range $1\,h^{-1}{\rm Mpc} < r < 10\,h^{-1}{\rm Mpc}$.

\section{Comparison with observations}
\label{Comparison}
In this section we present the predicted ACFs and their comparison with ACFs measured for observed galaxies. Then, we analyse the clustering properties of the LBGs.

%
%
\subsection{Observed ACF}
\label{Observed_ACF}
\cite{Rob2014} measured the ACF of LBGs at $z\sim3.8$ - $7.2$. They used the LBG samples of  \cite{Bouwens2015}. As listed in Table~\ref{Table:Flux_limit}, observations were carried out with eight different survey areas and flux limits. 

\cite{Rob2014} measured ACFs in each field. Owing to the finite survey area, the measured ACF is underestimated by a constant known as the integral constraint (IC):
\begin{equation}\label{w_obs}
w_{\rm true}(\theta)=w_{\rm obs}(\theta)+{\rm IC}.
\end{equation}
Because the IC value depends only on the size and shape of the survey area when the correlation slope, $\beta$, is fixed, the ${\rm IC}$ value is can be estimated using random catalogues generated on a field, which has the same size and shape as the survey area (see e.g. \citealp{Lee2006}). We note that \cite{Rob2014} fixed $\beta=0.6$ following previous studies \citep{Lee2006,Oberzier2006}. Errors in the measured ACF are estimated using bootstrap resampling as described in \cite{Ling1986}.

\cite{Rob2014} also obtained a single measurement of the ACF, combining the independently measured ACFs from the individual survey fields. Note that we do not attempt to reproduce the combined ACF. However, we use this measurement as a reference in cases where the measured ACF from the individual survey field has a large uncertainty due to a low number of LBGs.

\begin{figure*}
\begin{center}
\includegraphics[width=18cm]{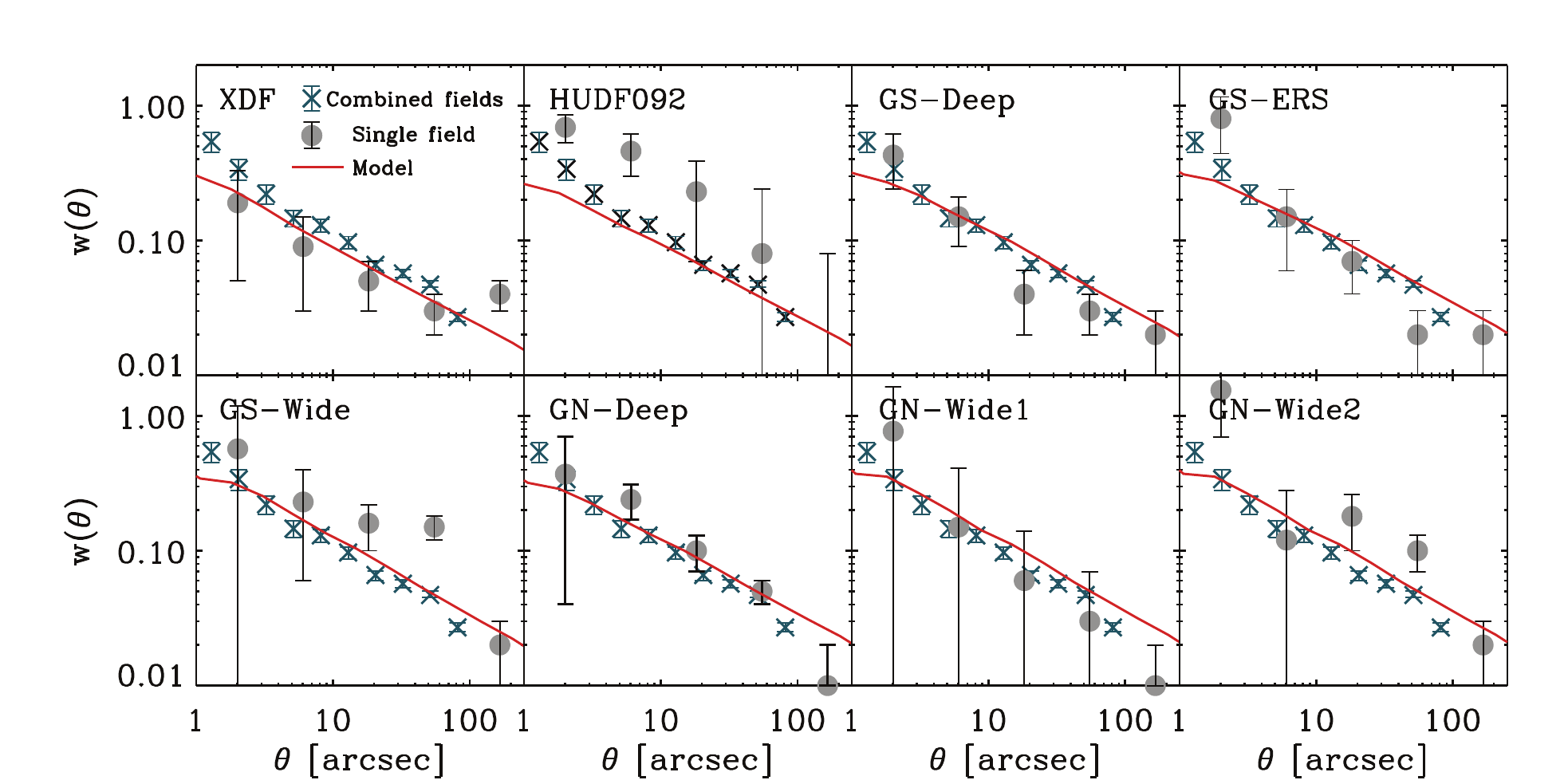}
\end{center}
\vspace{-3mm}
\caption{
The predicted angular clustering of LBGs using Obs-LBGs, shown as a solid line. The name of the field is labelled on each panel. Filled circles with error bars show the observed ACF measured from the individual field. The crosses with error bars show the observed combined ACF and are reproduced in each panel for reference. All errors are $1\,\sigma$ and estimated using bootstrap resampling \protect\citep{Ling1986}.
}
\label{Fig4}
\end{figure*}

%
%
\subsection{The comparison of the predicted ACF with observational measurement}
\label{Comparisons_with_observations}
In Fig.~\ref{Fig4}, we show the predicted and measured ACFs for the flux limits corresponding to each field. We also show the combined ACF for comparison. Note that the measured ACFs from individual fields are displayed using fewer bins than are used for the combined ACF to reduce the errors due to small numbers of objects. In the remainder of the paper, we investigate the clustering properties using the Obs-LBGs mock galaxy sample. To investigate the effect of the photometric scatter on the ACF, we revisit the differences between Intrinsic-LBGs and Obs-LBGs in Appendix~\ref{Effect_from_uncertainties}

All predicted ACFs are consistent with the measured ACFs for individual fields within $2\,\sigma$ errors, and are consistent with the combined ACF within $3\,\sigma$.
Overall, the predicted ACFs are in good agreement with both the measured ACFs for individual fields and the combined ACF. However, the predicted clustering amplitudes on small scales are lower than the combined ACF. Compared with the measured ACFs, the differences in the predicted clustering amplitude between fields are small. This is because the predicted ACFs are less affected by sample variance since the volume of the simulation is larger than that probed by the observations. Furthermore, since we predict the ACFs for each field using the same simulation box, the predicted ACFs are not subject to field-to-field variations, which are significant in typical HST observations \citep{Trenti2008}.

In general, the measured ACFs show similar shapes to the combined ACF. However, the ACFs measured from the two deep fields (XDF and HUDF092) show a different behaviour. The clustering amplitude from the XDF is lower than the amplitude from other fields, and is consistent with the predicted ACFs for the two deep fields. On the other hand, the measured ACF from HUDF092 shows the highest clustering amplitude. We interpret this as being caused by field-to-field variations. \cite{Bouwens2015} found field-to-field variations in the surface density in HUDF fields such that galaxies at $z\sim 4$ are relatively underdense.

%
%
\subsection{Dependance of clustering on luminosity}
\label{Clustering_dependence}

In the local Universe, the galaxy clustering strength is known to depend on luminosity \citep{Norberg2001,Zehavi2002}. Similarly, galaxy clustering at high redshifts ($z\gtrsim3$) has been shown to depend on rest-frame UV luminosity \citep{Ouchi2004,Ouchi2005,Cooray2006,Lee2006,Kashikawa2006b,Hildebrandt2009}. These observational results suggest a relation between luminosity and dark matter halo mass, with more massive haloes hosting brighter galaxies \citep[e.g., ][]{G&D2001}. This follows because massive haloes cluster more strongly than less massive haloes \citep[e.g., ][]{MW96}. Here, we investigate the predicted dependence of galaxy clustering on luminosity, considering a flux limit corresponding to the XDF. 

\begin{figure}
\begin{center}
\includegraphics[width=8.6cm]{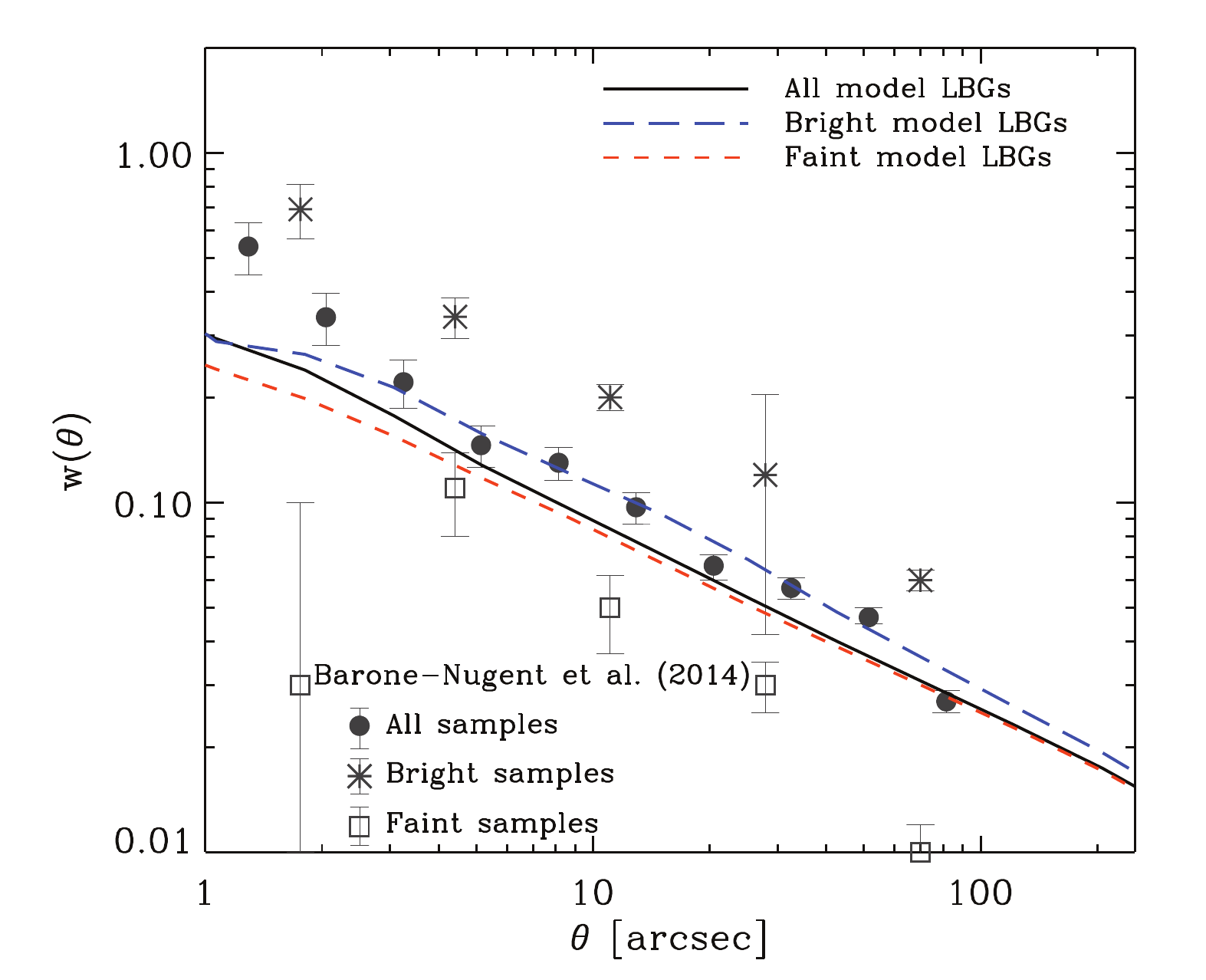}
\end{center}
\vspace{-3mm}
\caption{The predicted ACFs for bright and faint model LBGs, split using the rest-frame AB magnitude of -18.5. Solid (black), long-dashed (blue) and dashed (red) lines represent all, bright and faint LBGs, respectively. Filled circles (black), diamonds (dark grey) and squares (light grey) show, respectively, the measured ACFs for total, bright and faint LBGs from \protect\cite{Rob2014}.
}
\label{Fig5}
\end{figure}

\begin{figure*}
\begin{center}
\includegraphics[width=8.6cm]{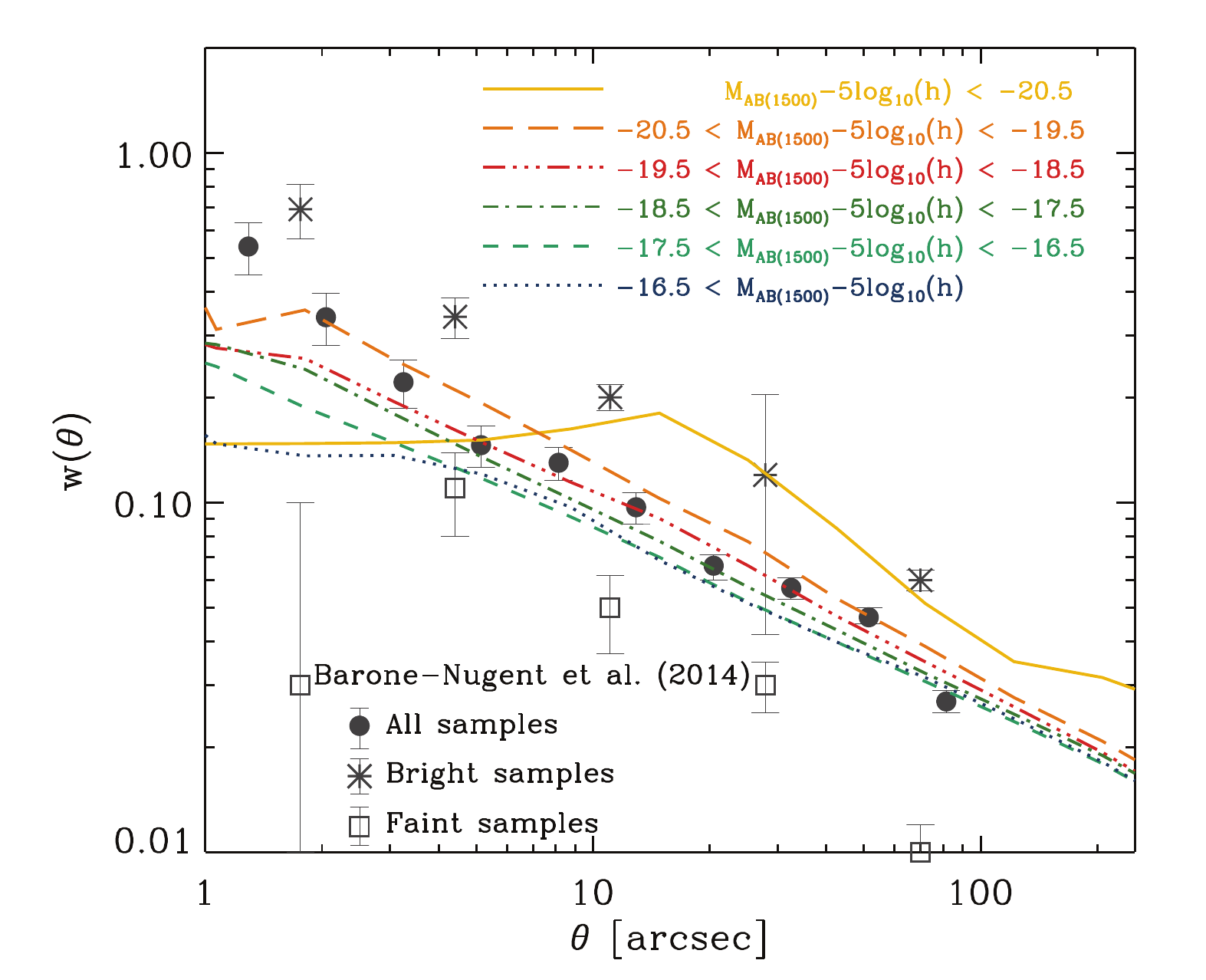}
\includegraphics[width=8.6cm]{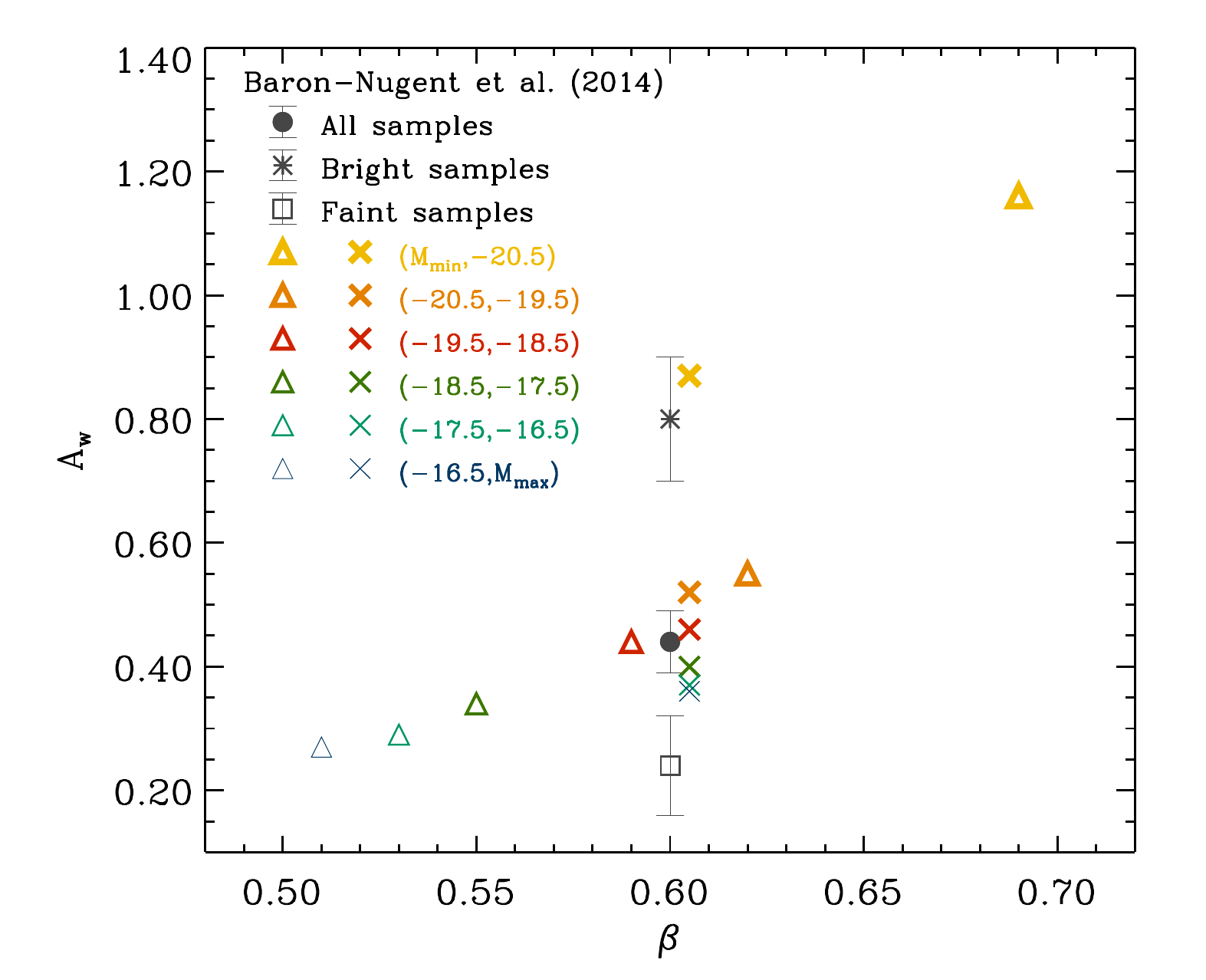}
\end{center}
\vspace{-3mm}
\caption{Left: the ACFs predicted in different magnitude bins as labelled. Right: the best fitting power-law parameters for the different magnitude bins. Crosses represent the best fitting $A_w$ values assuming a fixed value of $\beta$ as 0.6, and triangles represent the best fitting power-law parameters, as is assumed in the observational estimation. The legend indicates the different magnitude bins. The brightest magnitude bin  Filled circles (black), diamonds (dark grey) and squares (light grey) show the measured ACFs for total, bright and faint LBGs, respectively, from \protect\cite{Rob2014}.
}
\label{Fig6}
\end{figure*}

\cite{Rob2014} divided the LBG samples into bright and faint samples using a median rest-frame AB magnitude of -18.5 at $1700{\rm \AA}$. The measured ACFs are in agreement with previous results (listed above) that brighter galaxies are more strongly clustered than fainter ones. We divide the model LBGs into bright and faint subsets using the same magnitude cut. We note that for the far-UV luminosity, \cite{Rob2014} use the effective rest-frame wavelength of $1700\,{\rm \AA}$, while the GALFORM predictions we use correspond to the magnitude at a rest-frame wavelength $1500\,{\rm \AA}$. In principle, we need to correct for this difference. However, the difference is very small because the spectra of LBGs are nearly flat in $f(\nu)$ in this wavelength range \citep{Lacey2011}.
 
The combined ACF is somewhat complicated to interpret in terms of the clustering dependence on luminosity, since \cite{Rob2014} combined survey fields with different flux limits. We therefore show the model predictions for the XDF field, which contains the deepest observation. Fig.~\ref{Fig5} shows the predicted ACFs of all, bright and faint LBG samples. The predicted ACFs show a dependence on luminosity. However this is weaker than the measured one.

To analyse the dependence of clustering on luminosity, we show the predicted ACFs as a function of rest-frame UV magnitude in the left panel of Fig.~\ref{Fig6}. For comparison, we show the measured ACFs from \cite{Rob2014} for bright and faint LBG subsamples. Generally, we find that the model predicts that brighter LBGs have a higher clustering amplitude than fainter ones. The clustering amplitude in the brightest bin ($M_{\rm AB(1500)}-5{\rm log}(h) < -20.5$) is comparable with that of the observed bright samples. In the other magnitude bins, the clustering amplitudes also a show dependence on luminosity, but this trend is not as strong as observed. Previous studies also reported that the ACFs predicted from semi-analytic simulations show a weaker clustering dependence on luminosity compared with that measured from observations \citep{Wechsler2001,Kashikawa2006b}. The latest GALFORM model shows a similar discrepancy with observations for the dependence of clustering on optical luminosity in the local universe (see e.g. \citealp{Campbell2015}).

The right panel of Fig.~\ref{Fig6} shows the best fitting power-law parameters for the correlation function amplitude, $A_w$, and the correlation slope, $\beta$, as a function of rest-frame UV magnitude. Note that we find the best fitting values by considering only angular separations larger than $10$ arcsec. We plot the measured $A_w$ from all samples, and from the bright and faint subsamples from \cite{Rob2014} for comparison. Note that \cite{Rob2014} fixed the value of $\beta$ to 0.6. We firstly compare the predicted $A_w$ values assuming $\beta=0.6$ to those measured from the observations. The value in the brightest bin ($M_{\rm AB(1500)}-5{\rm log}(h)<-20.5$) is comparable to the value measured for the bright sample. The values in the other bins ($-20.5 < M_{\rm AB(1500)}-5{\rm log}(h)$) are all comparable with the the measured value for the combined LBG sample. Overall, the predicted $A_w$ depends weakly on the luminosity except at the highest luminosities.

When we allow the value of $\beta$ to vary, $A_w$ shows a stronger dependence on luminosity. We also find that the values of $\beta$ gradually increase with luminosity as well as the value of $A_{w}$. This result is consistent with the measurement from \cite{Kashikawa2006b}. 

Interestingly, for the ACFs measured from observations, we find that the clustering amplitude of faint samples decreases at small angular separations ($\theta \lesssim 5''$). This decrease is not seen for the bright samples. We interpret this as showing that massive haloes host multiple bright LBGs, so there is a contribution from central-satellite galaxy pairs, while faint galaxies tend not to be satellites \citep{Kashikawa2006b}. 

This finding is in contrast to the model prediction for the clustering amplitude in the two brightest bins ($M_{\rm AB(1500)}-5{\rm log}(h) < -20.5$ and $-20.5<M_{\rm AB(1500)}-5{\rm log}(h)<-19.5$), which show a decrease at small angular separations. We checked that changing the magnitude range does not alter this trend.
 
\begin{figure*}
\begin{center}
\includegraphics[width=17cm]{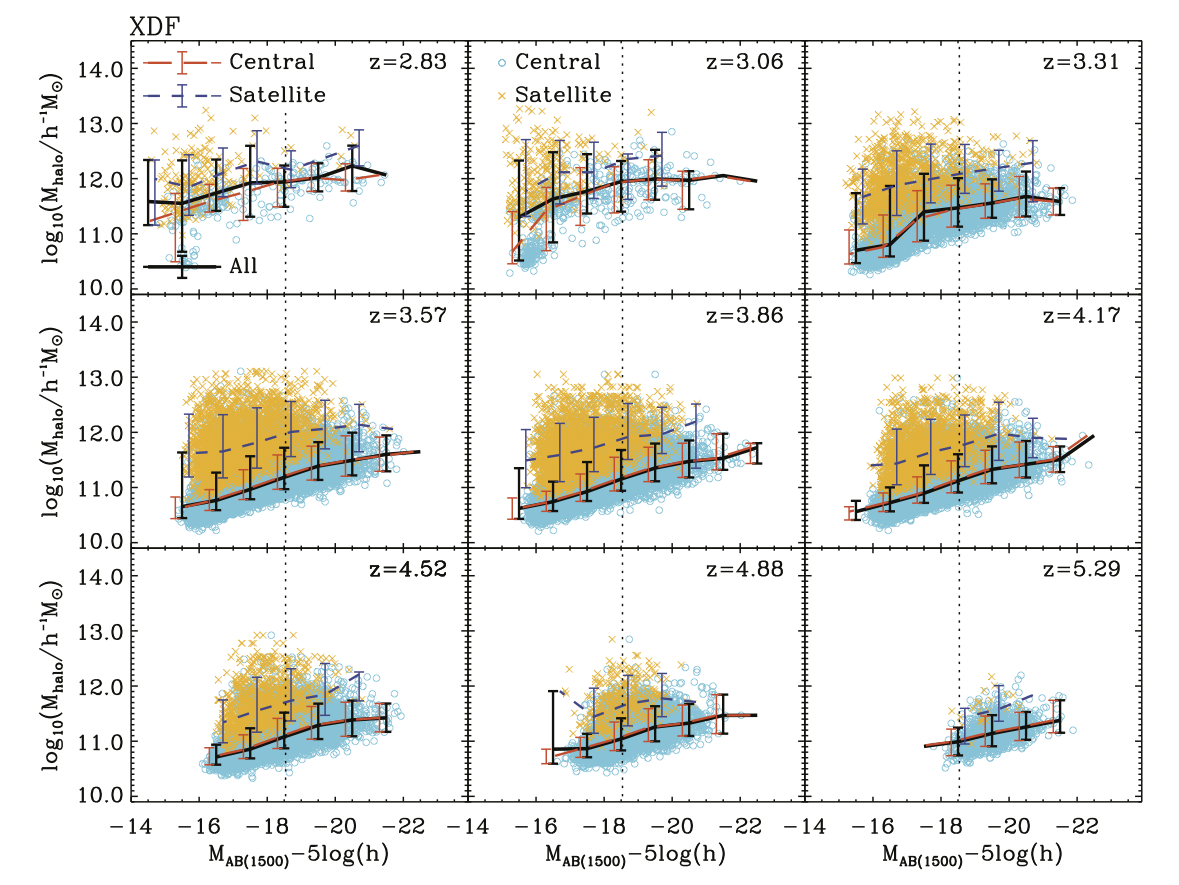}
\end{center}
\vspace{-3mm}
\caption{Predicted relation between UV luminosity and host halo masses of LBGs selected in galaxy catalogues at different redshifts as labelled. Here, we use the flux limits for the XDF field. Open circles (cyan) and crosses (yellow) show central and satellite LBGs, respectively. Solid (black), long-dashed (red) and dashed (blue) lines represent host halo masses of all, central and satellite LBGs, respectively. Vertical bars represent 10-90 percentile ranges for the mass. In each panel, vertical dotted lines indicate a median magnitude of $M_{{\rm AB}(1500)}-5{\rm log(h)}=-18.54$ as used by \protect\cite{Rob2014} to distinguish between faint and bright LBGs.
}
\label{Fig7}
\end{figure*}

We can explain the small scale clustering by considering the number of satellite LBGs. In Fig.~\ref{Fig7}, we show the relation between the host halo masses and the rest-frame UV luminosity of the model LBGs. We also show the median value of the host halo mass as a function of luminosity with 10-90 percentile ranges for all LBGs, and for central and satellite LBGs. The median values for all LBGs are almost identical to those for central LBGs. The 10-90 percentile ranges for all LBGs are also consistent with the ranges for central LBGs, but the ranges for all LBGs broaden toward massive halo masses at faint luminosities. This behaviour is because most bright LBGs are central galaxies. The clustering amplitude of bright LBGs is dominated by central-central LBG pairs (i.e. the contribution from the two-halo term). 

In practice, the two brightest bins ($M_{\rm AB(1500)}-5{\rm log}(h) < -20.5$ and $-20.5<M_{\rm AB(1500)}-5{\rm log}(h)<-19.5$) include only 2.0 and 3.9 per cent satellite LBGs, respectively, while the two faintest bins ($-16.5 < M_{\rm AB(1500)}-5{\rm log}(h)$ and $-17.5<M_{\rm AB(1500)}-5{\rm log}(h)<-16.5$) contain 11.5 and 7.8 per cent satellite LBGs. For this reason the clustering amplitude of bright LBGs decreases at small angular separations. This implies that the central-satellite LBG pairs play an important role in shaping the amplitude of the ACF on small scales, and that the model predicts fewer satellite LBGs than is inferred from the ACF measured from observations.
As mentioned above, the model ACFs show the opposite trend to the observations for the clustering amplitude at small angular separations. This implies that faint samples of real galaxies contain more central galaxies than the model. However, this conclusion is tentative because the uncertainties in the measurement of ACF are larger than those presented in Fig.~\ref{Fig6}, if the uncertainties due to cosmic variance and the use of a fixed $\beta$ are included.

In Fig.~\ref{Fig8}, we show the median host halo mass in bins of the rest-frame UV magnitude as a function of redshift for Obs-LBGs assuming the detection limits of the XDF field. We find that central LBGs predominantly reside in $\sim 10^{11}-10^{12}h^{-1}M_{\rm \odot}$ haloes, and that satellites reside in $\sim 10^{12}-10^{13}h^{-1}M_{\rm \odot}$.

%
%
%
%
\subsection{Halo Occupation Distribution}
\label{HOD}
\begin{figure}
\begin{center}
\includegraphics[width=8.5cm]{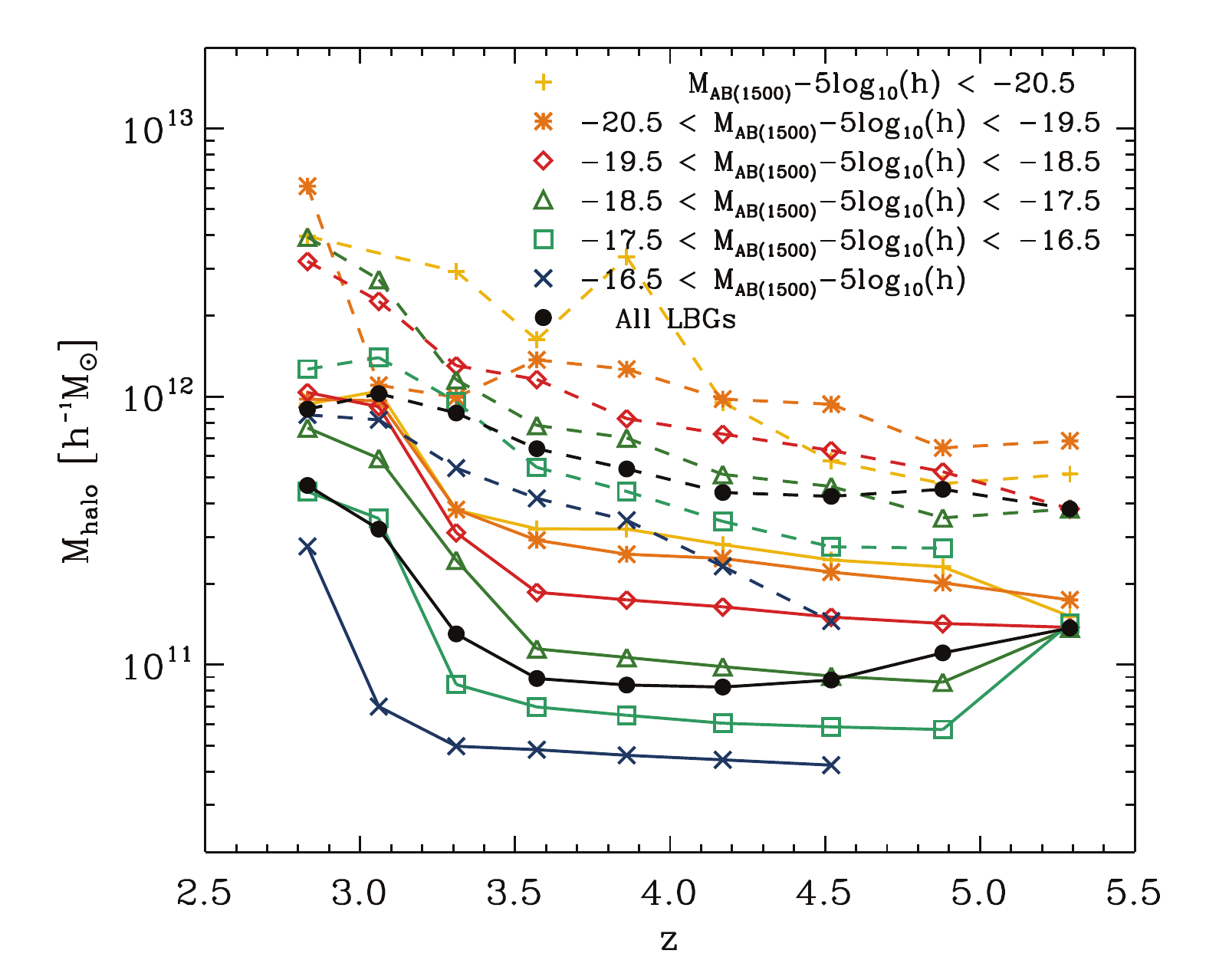}
\end{center}
\vspace{-3mm}
\caption{
Median host halo mass of model Obs-LBGs as a function of redshift. Solid and dashed lines represent central and satellite galaxies, respectively. Filled circles denote the median host halo mass of all LBGs and the legend indicates the rest-frame magnitude at $1500{\rm \AA}$.
}
\label{Fig8}
\end{figure}
To quantify the contribution of satellite and central galaxies to the clustering signal, the two-point correlation function can be decomposed into one- and two-halo terms \citep{Zheng2004},
\begin{equation}\label{w_decoupled}
\xi(r) = \xi^{\rm 1h}(r)+\xi^{\rm 2h}(r)+1.
\end{equation}
On scales larger than the virial radius of haloes, pairs consist of galaxies in separate haloes, producing a two-halo term. On small scales, pairs consist mainly of galaxies inside the same halo, producing a one-halo term. Thus, the two-halo term is due to central-central or central-satellite  galaxy pairs but in different haloes, and the one-halo term is due to central-satellite and satellite-satellite galaxy pairs from the same halo. 

In the Halo Occupation Distribution (HOD) model, the galaxy population can be split into centrals and satellites. The simplest model for the average number of the central HOD is \citep{Zheng2005}
\begin{equation}\label{HOD_cen}
\left<N_{\rm c}\right>_M =
\begin{cases}
1 & \text{for }M>M_{\rm min}, \\
0 & \text{otherwise.}
\end{cases}
\end{equation}
The satellite HOD can be written as \citep{Zheng2005}
\begin{equation}\label{HOD_sat}
\left<N_{\rm s}\right>_M = \left(\frac{M-M_{\rm cut}}{M_1} \right)^{\alpha},
\end{equation}
where $M_{\rm cut}$ is the minimum mass of haloes that host satellite galaxies.

\begin{figure*}
\begin{center}
\includegraphics[width=15cm]{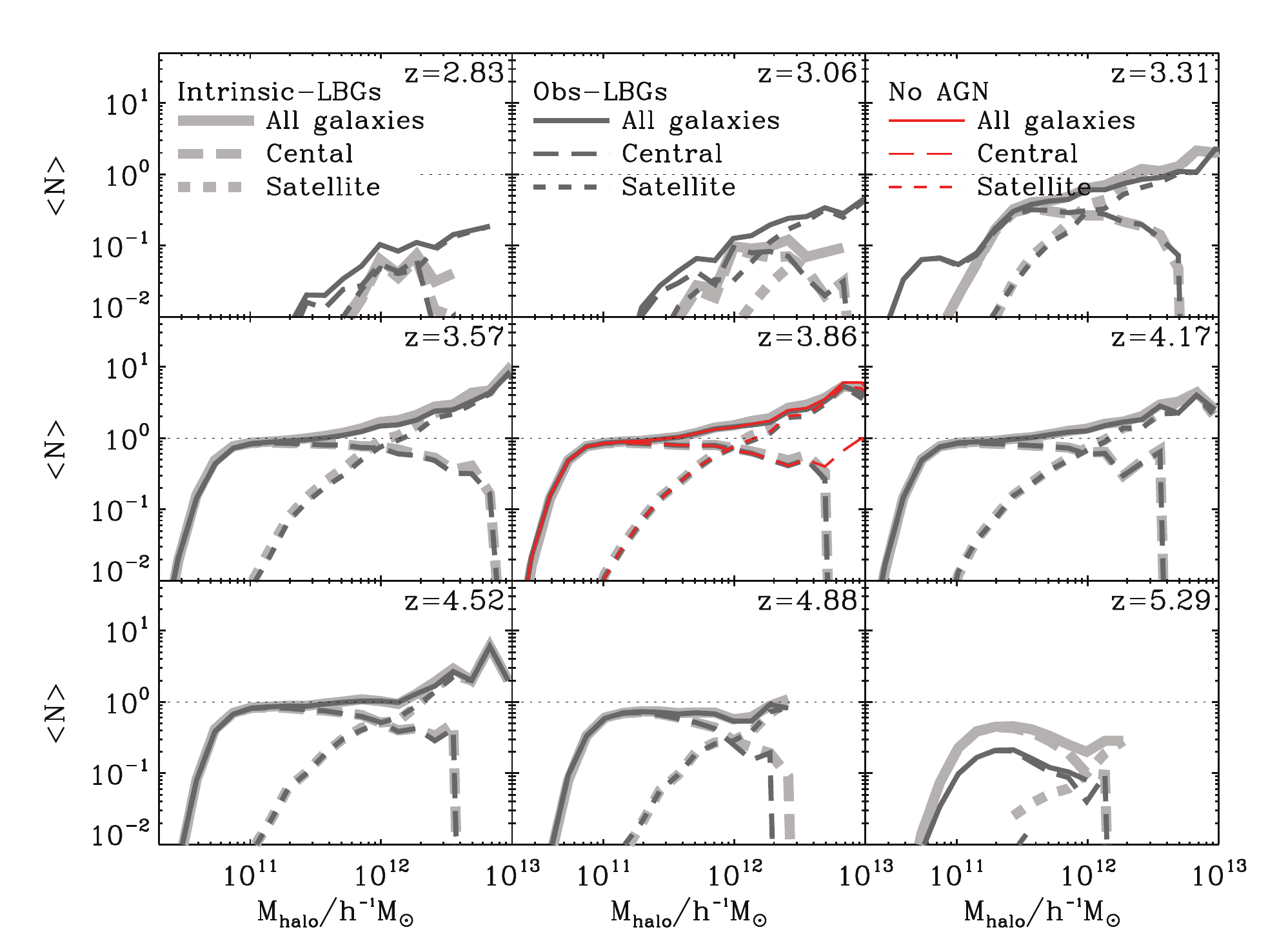}
\end{center}
\vspace{-3mm}
\caption{The predicted model HOD at different redshifts as labelled in each panel. Solid, long-dashed and dotted lines represent total, central and satellite LBGs, respectively. Thick (light grey) and modest (dark grey) lines represent, respectively, the predicted HOD using Obs-LBGs and Intrinsic-LBGs. Thin (red) lines at $z=3.86$ represent the predicted HOD when we switch off AGN feedback.
}
\label{Fig9}
\end{figure*}

Since the redshift distribution of observed LBGs is broad, the HOD derived from observations contains information for galaxies selected over a range of redshifts. Thus, we show HODs for galaxies selected in GALFORM over a range of redshifts. We show the HODs for Obs-LBGs in Fig.~\ref{Fig9}. At redshifts $3.5\lesssim z\lesssim4.5$, the HODs are very similar to one another. At the two lowest redshifts ($z=2.83$ and 3.06) and the highest redshift ($z=5.29$), the mean number of galaxies is less than unity at all halo masses. This is because the number of LBGs is very small due to the colour selection. The minimum halo mass which hosts an LBG detectable in the HST observations is $\sim 1.6\times 10^{10}\,h^{-1}M_{\rm \odot}$ in the GALFORM model at these redshifts. 

We find that the mean number of central galaxies with star formation rate sufficiently high for detection as a LBG predicted by GALFORM drops sharply in massive haloes ($M_{\rm halo} \gtrsim 5\times10^{12}\,h^{-1}M_{\rm \odot}$). This sudden drop is caused by AGN feedback, which suppresses star formation in massive haloes by shutting down gas cooling (see \citealp{Bower2006} for more details). For present day galaxies, the HOD of central galaxies selected by $b_{\rm J}$ band luminosity shows a drop above a halo mass of $\sim 10^{12}h^{-1}M_{\rm \odot}$ \citep{Kim2009}. To illustrate this, we plot the HOD for a model in which we switch off AGN feedback at $z=3.86$. As shown in the center panel of Fig.~\ref{Fig9}, the drop at high mass is not detected in the central HOD when we remove AGN feedback. Overall, this trend is consistent with the result that central LBGs are  predominantly in $\sim 10^{11}-10^{12}h^{-1}M_{\rm \odot}$ haloes, and satellite LBGs are in $\sim 10^{12}-10^{13}h^{-1}M_{\rm \odot}$ haloes (Fig.~\ref{Fig8}).

We find that the HOD for satellite LBGs also shows a cutoff in halo mass above a few $10^{13}\,h^{-1}M_{\rm \odot}$. However, this arises because the simulation volume does not contain haloes above this cutoff mass due to its finite volume and the low space density of massive haloes, so that we cannot predict the HOD robustly at these masses.

\begin{figure}
\begin{center}
\includegraphics[width=8.5cm]{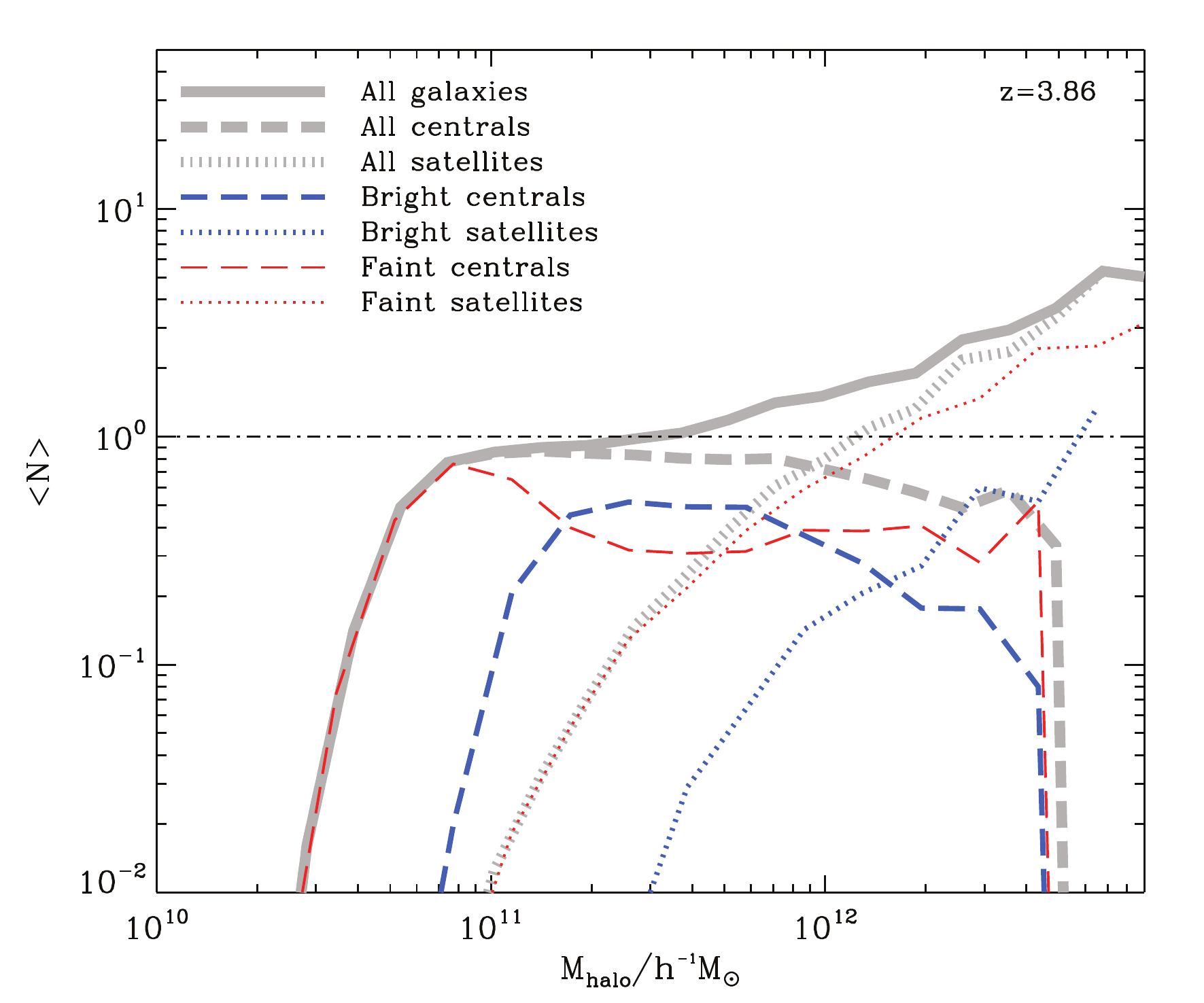}
\end{center}
\vspace{-3mm}
\caption{The same as Fig.~\ref{Fig9}, but at $z=3.86$. Thick (light grey) lines represent the predicted HOD of all galaxies. Modest (blue) and thin (red) lines represent bright and faint model LBGs, respectively, divided using the rest-frame AB magnitude of -18.5.
}
\label{Fig10}
\end{figure}

In \S.~\ref{Clustering_dependence} we argued that the model appears to predict fewer bright satellite LBGs than suggested by observations. Before concluding we investigate whether this can be seen in the HOD. In Fig.~\ref{Fig10} we plot the predicted HOD of bright and faint model LBGs, divided using the rest-frame AB magnitude of -18.5. The predicted HOD of faint satellite LBGs is found to be comparable to that for all LBGs. On the other hand, the predicted HOD of bright satellite LBGs is smaller, indicating that the model does not predict many bright satellite galaxies.

\section{Summary and Discussion}\label{Summary}
We have investigate the clustering properties of Lyman-break galaxies (LBGs) at $z\sim4$. Using the hierarchical galaxy formation model GALFORM, we predict the angular correlation function (ACF) of LBGs selected in the model. We compare the model predictions with the clustering measurements made by \cite{Rob2014}, who used combined survey fields consisting of the Hubble eXtreme Deep Field (XDF) and CANDELS.

To predict the angular clustering of LBGs we use, for the first time, a semi-analytical model which includes the effects of AGN feedback. In an earlier study of the nature of LBGs using GALFORM, \cite{Lacey2011} used the model of \cite{Baugh2005}, which does not include AGN feedback, to investigate their host halo mass and bias. \cite{Lacey2011} also considered the model of \cite{Bower2006}, which does include AGN feedback, but this model overpredicts the bright end of the rest-frame UV luminosity function and hence the host halo mass and bias were not presented for this case. Hence, this represents the first study of the clustering of LBGs in a model with AGN feedback which also reproduces the observed abundance of LBGs.

Prior to this paper, the most recent study using semi-analytical models to study the clustering of LBGs is that by \cite{Kashikawa2006b}, who used the model of \cite{Nagashima2005}. There are three key differences between the model of Nagashima et~al and the one used here. First, the model of \cite{Nagashima2005} did not include AGN feedback. Instead, in order to suppress the formation of bright  galaxies Nagashima et~al. restricted gas cooling by hand in haloes above a given circular velocity. Second, to calculate dust extinction which is critical to compute the photometric properties of LBGs, \cite{Nagashima2005} adopted a foreground screen dust model, whereas GALFORM computes the dust extinction more realistically, considering the stars and dust to be mixed together \citep[e.g.][]{Lacey2011}. Lastly, in the modelling of star formation, \cite{Nagashima2005} considered only the total cold gas mass, whereas the model of \cite{Lagos2012} uses an impoved star formation treatment which splits the interstellar medium into atomic and molecular hydrogen, with only the latter taking part in star formation.

In another recent study, \cite{Jose2013} predicted the angular clustering of LBGs using the simple model introduced by \cite{Samui2009} and compared with observations. To generate the formation histories of dark matter halos \cite{Samui2009} started from an analytical form of the halo mass function rather than an N-body simulation and they mainly focused on one process, the star formation rate. Furthermore, \cite{Jose2013} applied the {\it same} dust attenuation to all model galaxies and did not attempt to reproduce galaxies at the present day.

Ours is the first study to compare model predictions with the new clustering measurements of LBGs at $z\sim4$. Since \cite{Rob2014} measured the angular clustering in each survey field independently, this measurement gives us an estimate of sample variance by comparing the results from different fields. In addition, the measurement from XDF field allows us to investigate the clustering of the fainter LBGs. 

We confirm the conclusion reached using independent models that bright galaxies in the model are more strongly clustered than faint ones. This dependence of clustering on luminosity is in qualitative agreement with the observations, but is weaker than is inferred from clustering measurements. We find that for bright LBGs the clustering amplitudes at small angular separations are predicted to be lower than estimated from the observations. This may imply that the number of bright satellite LBGs is smaller in the model than in the real Universe. Although we find discrepancies, two factors in the observation must be considered. First, the measurement presented in \cite{Rob2014} is somewhat complicated to interpret, because they combined the observed samples from survey areas with different flux limits. Second, if we include the uncertainties associated with cosmic variance and assuming a fixed $\beta$, the uncertainties in the measured ACFs will become larger than those presented. Therefore, larger surveys of observed LBGs are needed. In addition, to further investigate the discrepancies, we need to compare the predicted clustering with different models using various galaxy formation physics in future work.

We have investigated the effect of the photometric scatter on the ACF. We find that the predicted ACF using Obs-LBGs shows lower clustering amplitude than that predicted using Intrinsic-LBGs without including the photometric scatter. This effect is larger for faint LBGs, because of the influx of the galaxies in low mass haloes into the sample. This trend could amplify the dependence of clustering on luminosity. In practice, the clustering dependence on luminosity of Obs-LBGs is stronger than that of Intrinsic-LBGs, although the dependence is still weaker than observations. We also find the predicted clustering amplitude using Obs-LBGs decreases more on small scales than on large scales. This may affect the interpretation of clustering for the one-halo term. Thus, future analyses need to pay attention to the photometric scatter when comparing the properties of model LBGs with those of observational LBGs.

We have analysed the HOD of $z\sim4$ LBGs. We find that central LBGs predominantly reside in $\sim 10^{11}-10^{12}h^{-1}M_{\rm \odot}$ haloes and satellites reside in $\sim 10^{12}-10^{13}h^{-1}M_{\rm \odot}$ for the detection limit of XDF field. We also find that the mean number of central LBGs drops sharply in massive haloes mass, due to AGN feedback. However, the drop in central galaxies is swamped by the satellite HOD which populates the larger haloes. This effect of AGN feedback is not normally considered in empirical HOD modelling for LBGs (e.g. \citealp{Hamana2004,Cooray2006,Lee2006,Hildebrandt2009}.)

\vspace{5mm}
{\bf Acknowledgments} We would like to thank Masami Ouchi, Yuichi Harikane and Jae-Woo Kim for useful comments and discussions. HSK is supported by a Discovery Early Career Researcher Awards (DE140100940) from the Australian Research Council. JSBW acknowledges the support of an Australian Research Council Laureate Fellowship. CMB acknowledge receipt of a Research Fellowship from the Leverhulme Trust. This work was supported in part by the Science and Technology Facilities
Council consolidated grant ST/L00075X/1 to the ICC. The Millennium II Simulation was
carried out by the Virgo Consortium at the supercomputer centre of the
Max Planck Society in Garching.
Calculations for this paper were partly performed on the ICC Cosmology Machine, which is part of the DiRAC Facility jointly funded by ST/K0032671, the Large Facilities Capital Fund of BIS,
and Durham University.  
\newcommand{\noopsort}[1]{}

\bibliographystyle{mn2e}

\bibliography{Cross}

%
%
\appendix

\section{Effect on the angular clustering from observational uncertainties}
\label{Effect_from_uncertainties}
In Section~\ref{LBG_select} we described how we select model LBGs. We use two different model LBGs, the Intrinsic-LBGs, which are selected using the intrinsic colours, i.e. no photometric scatter, and the Obs-LBGs, which are selected using colours that include the photometric scatter. Here, we investigate how this photometric scatter affects the angular clustering.

\begin{figure*}
\begin{center}
\includegraphics[width=18cm]{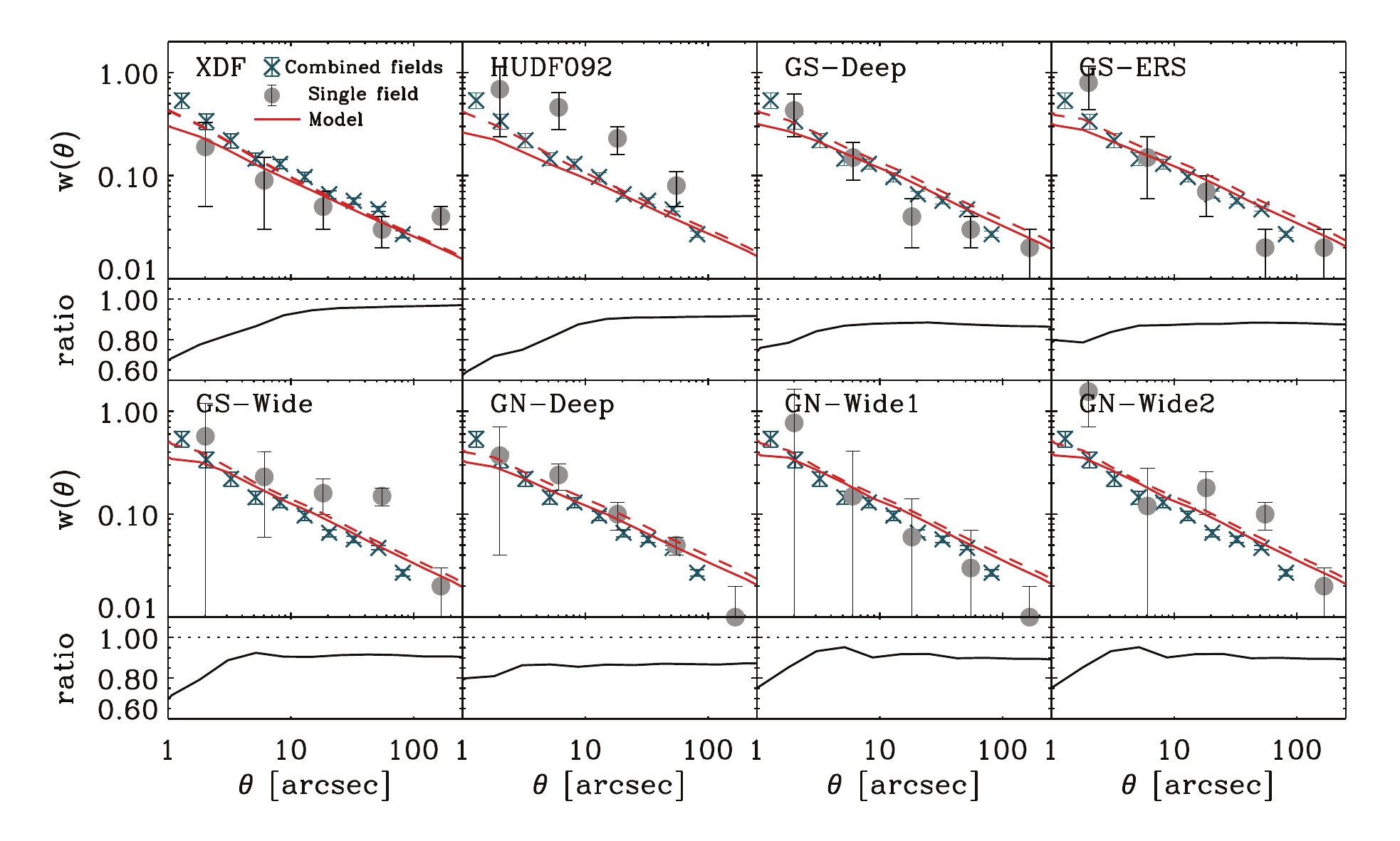}
\end{center}
\vspace{-3mm}
\caption{The same as Fig.~\ref{Fig4} but we plot the ACF using Intrinsic-LBGs (dashed line) as well as Obs-LBGs (solid line). In each panel the bottom sub-panels show the ratio of the ACF from Obs-selection to the ACF from Intrinsic-selection.
}
\label{FigA1}
\end{figure*}

Fig.~\ref{FigA1} shows the predicted ACFs using LBGs selected using the two treatments of galaxy magnitude for the different flux limits corresponding to each field. We plot the ratio of the predicted ACF using Obs-LBGs to the predicted ACF using Intrinsic-LBGs. We find that the predicted clustering amplitude using Obs-LBGs is lower in amplitude for all survey fields.

Contamination by foreground galaxies reduces the clustering amplitude \citep{Ouchi2004, Kashikawa2006b}. The contamination is defined as being due to galaxies located below the boundary redshift, where the boundary redshift is $z=3.5$ in the case of LBGs at $z\sim4$ \citep{Yoshida2006}. The correlation function amplitude, $A_w$, is decreased by a factor of $(1-f_c)^2$, where $f_c$ is the contamination rate \citep{Ouchi2004, Kashikawa2006b}. In practice, the contamination rate of the Obs-LBGs is larger than that of Intrinsic-LBGs by a factor of 2. Most of the contamination is at very low redshift ($0.2 \leq z \leq 0.8$) which we do not use in our calculations, although the rest of the galaxies are close to the boundary redshift \citep{Yoshida2006}. To quantify how the contamination near the boundary redshift affects the clustering amplitude requires a detailed study that is beyond the scope of this paper.

\begin{figure}
\begin{center}
\includegraphics[width=8.6cm]{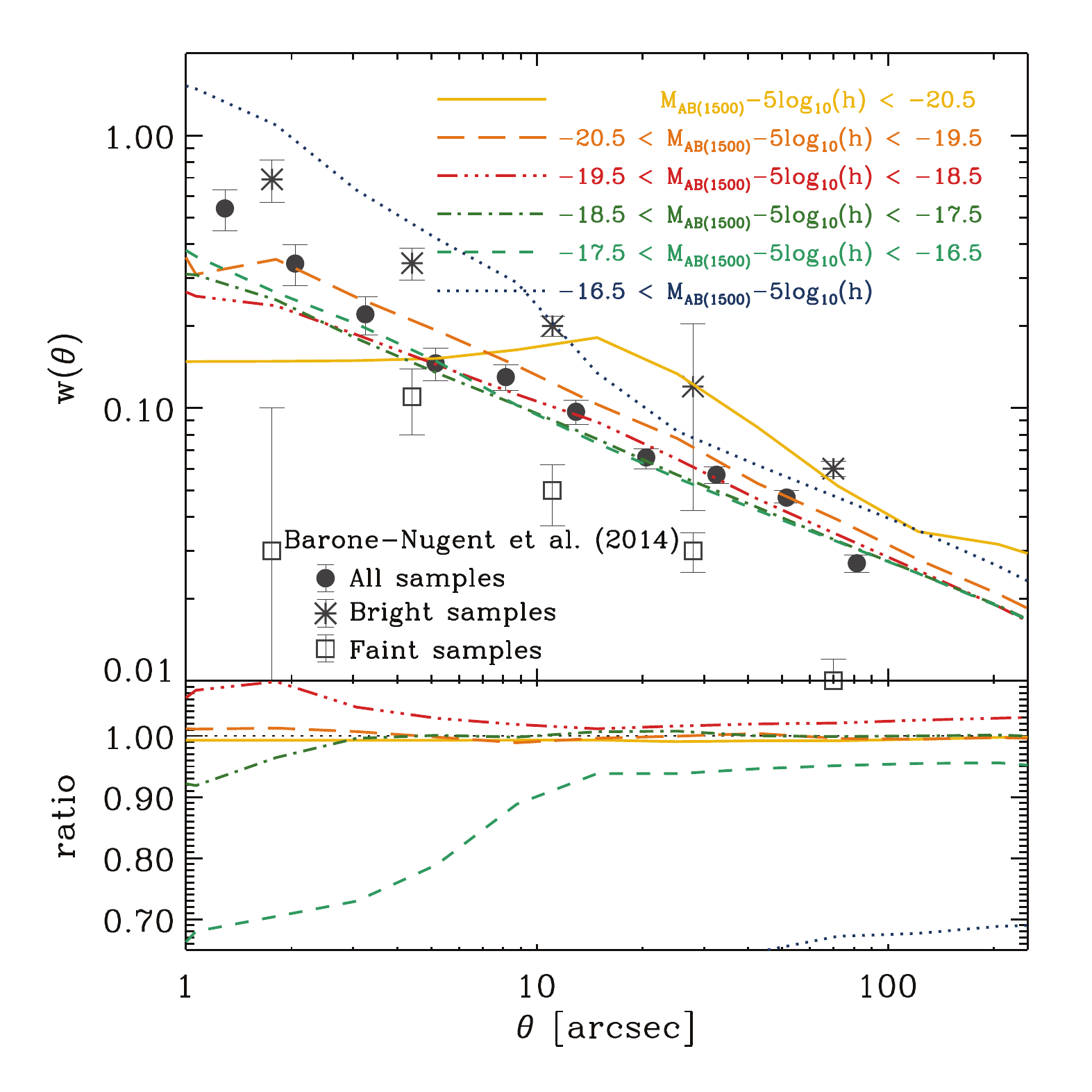}
\end{center}
\vspace{-3mm}
\caption{The same as the left panel of Fig.~\ref{Fig6} but we plot the ACF using Intrinsic-LBGs. Bottom sub-panel shows the ratio of the ACF using Obs-LBGs to the ACF using Intrinsic-LBGs.
}
\label{FigA2}
\end{figure}

Another possible factor which reduces the clustering amplitude is photometric scattering of galaxies into the sample from low mass haloes. Because of the photometric scatter, galaxies can have a brighter observed magnitude than their intrinsic magnitude and vice versa. As seen in Fig.~\ref{Fig7}, the rest-frame UV magnitude is proportional to the host halo mass. Selected galaxies can therefore reside in lower mass haloes than other galaxies having the same luminosity, and contribute to a reduced galaxy bias. Fig.~\ref{FigA2} shows the predicted ACF using Intrinsic-LBGs as a function of the rest-frame UV magnitude and the ratio of the ACF using Obs-LBGs to the ACF using Intrinsic-LBGs. As expected, the clustering amplitudes using Obs-LBGs in the faint bins are reduced more than those in bright bins. For this reason, unless model galaxies include the photometric scatter, the dependence of clustering on luminosity could be underestimated. We find that the predicted ACFs using Intrinsic-LBGs do not show a dependence on luminosity in faint bins (Fig.~\ref{FigA2}), but that the predicted ACFs using Obs-LBGs do show a weak dependence (Fig.~\ref{Fig6}). 

\begin{figure}
\begin{center}
\includegraphics[width=8.5cm]{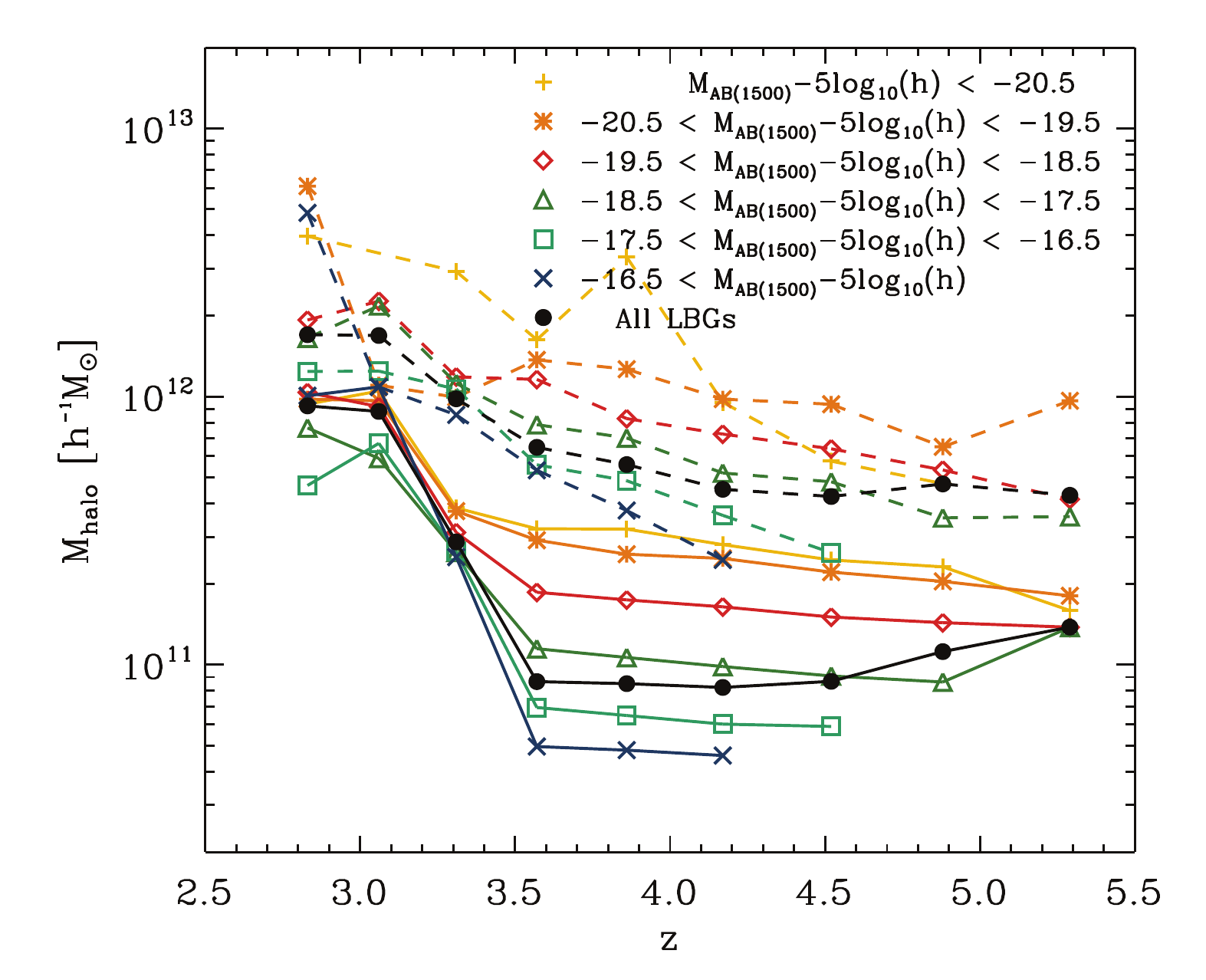}
\end{center}
\vspace{-3mm}
\caption{
Median host halo mass of model for Intrinsic-LBGs as a function of redshift. Solid and dashed lines represent central and satellite galaxies, respectively. Filled circles denote the median host halo mass of all LBGs and the legend indicates the rest-frame magnitude of $1500{\rm \AA}$.
}
\label{FigA3}
\end{figure}

In Fig.~\ref{FigA3}, we show the median host halo mass in the same rest-frame UV magnitude bins as a function of redshift for Intrinsic-LBGs. The figure shows that median host halo masses for Obs-LBGs are lower than those for Intrinsic-LBGs, especially in faint bins and low redshifts. We also find that central LBGs predominantly reside in $\sim 10^{11}-10^{12}h^{-1}M_{\rm \odot}$ haloes, and that satellites reside in $\sim 10^{12}-10^{13}h^{-1}M_{\rm \odot}$ in both cases.

We find that the clustering amplitude on small scales decreases more using Obs-LBGs. We explain this change using the fact that colours of satellites are statistically more likely to be scattered into the sample than those of centrals, since satellites are generally fainter than centrals. We find that the colour distribution gradually moves down with decreasing redshift (Fig.~\ref{Fig1}). Near the boundary redshift ($z\sim3.5$), the satellites that are located inside the colour selection region but located close to the lower boundary of the region are more likely to deviate from the region than centrals. Consequently, the fraction of satellites among Obs-LBGs is lower than that of Intrinsic-LBGs. For this reason, the clustering amplitude using Obs-LBGs decreases on small scales compared with the amplitude using Intrinsic-LBGs. We also checked that deeper magnitude limits produce larger amplitude change on small scales. This is because the fraction of satellites increases when we consider deeper magnitude limits, although the photometric scatter decrease. 

The clustering of LBGs from observations shows an enhanced clustering amplitude compared to a power law on small scales \citep{Ouchi2005,Lee2006,Cooray2006,Kashikawa2006b,McLure2009,Hildebrandt2009}. The predicted ACF using Intrinsic-LBGs shows this trend, but the ACF using Obs-LBGs does not show this due to the contribution of photometric errors. Although the predicted ACF using Intrinsic- and Obs-LBGs are both comparable with the measured ACF from observations within $3\,\sigma$ errors, this reduced amplitude on small angular scales may affect the interpretation of the one-halo term. Therefore, we emphasise again that inclusion of the photometric scatter in model galaxies is important to compare their properties with observations.

We find that the HODs for Intrinsic-LBGs and for Obs-LBGs are similar to one another. The only differences appear at redshifts which deviate from the targeted redshift range. At low redshift, the HODs for Obs-LBGs have a smaller minimum mass ($M_{\rm min}$). Thus, the fact that smaller $M_{\rm min}$ produces a smaller clustering amplitude, especially on small scales, is consistent with the result shown in Fig.~\ref{FigA1}.

\label{lastpage}
\end{document}